\documentclass[preprint]{JHEP} % 10pt is ignored!
\usepackage{amsmath}
\usepackage{amssymb,amsfonts}
\usepackage{epsfig}

\newcommand\fverb{\setbox\pippobox=\hbox\bgroup\verb}
\newcommand\fverbdo{\egroup\medskip\noindent%
            \fbox{\unhbox\pippobox}\ }
\newcommand\fverbit{\egroup\item[\fbox{\unhbox\pippobox}]}
\newbox\pippobox

\newcommand{\vecto}[2]{\left( \begin{array}{c} #1 \\ #2 \end{array} \right) }
\newcommand{\matrto}[4]{\left( \begin{array}{cc} #1 & #2 \\ #3 & #4 \end{array} \right) }

\newcommand{\nn}{\nonumber}

\newcommand{\C}{{\cal C}}
\newcommand{\E}{{\cal E}}

\newcommand{\old}[1]{#1_{\mathrm{old}}}
\newcommand{\oldp}[1]{(#1)_{\mathrm{old}}}
\def\zz{z}

\newcommand{\be}{\begin{equation}}
\newcommand{\ee}{\end{equation}}
\newcommand{\ben}{\begin{enumerate}}
\newcommand{\een}{\end{enumerate}}
\renewcommand{\sp}{\ ,\qquad}
\makeatletter
\renewcommand{\@makefnmark}{\mbox{$^{\ddagger\@thefnmark}$}}
\renewcommand{\subsection}{\@startsection
  {subsection}{2}{0pt%-1em
}{-\baselineskip}{0.5\baselineskip}
  {\normalfont\normalsize\itshape}}
\makeatother
\numberwithin{table}{section}

\newcommand{\gym}{g_{\mathrm{YM}}}
\newcommand{\geff}{g_{\mathrm{eff}}}
\newcommand{\gseff}{g_s^{\mathrm{eff}}}
\newcommand{\aeff}{a^{\mathrm{eff}}}
\newcommand{\Ord}{{\cal{O}}}
\newcommand{\xnc}{\zz_{\mathrm{nc}}}
\newcommand{\xt}{\zz_{\mathrm{t}}}
\newcommand{\hes}{\mbox{Hes}}

\title{Phase Structure of Non-Commutative Field Theories and Spinning Brane Bound States}

\author{T. Harmark and N.A. Obers\thanks{
Work supported in part by TMR network ERBFMRXCT96-0045.}\\
    Niels Bohr Institute and Nordita, Blegdamsvej 17, DK-2100 Copenhagen, Denmark \\
    E-mail: \email{harmark@nbi.dk}, \email{obers@nordita.dk} }
\preprint{ NBI-HE-99-47  \\
NORDITA-1999/75 HE \\
\hepth{9911169}}    % OR: \preprint{Aaaa/Mm/Yy\\Aaa-aa/Nnnnnn}
\abstract{General spinning brane bound states are constructed,
along with their near-horizon limits which are relevant as dual
descriptions of non-commutative field theories. For the spinning
D-brane world volume theories with a $B$-field a general analysis
of the gauge coupling phase structure is given, exhibiting various
novel features, already at the level of zero angular momenta.
We show that the thermodynamics is equivalent to the commutative case at 
large $N$  
and we discuss the possibility and consequences of finite $N$.
As an application of the general analysis, the
range of validity of the thermodynamics for the NCSYM is
discussed. In view of the recently conjectured existence of a 
7-dimensional NCSYM, the thermodynamics of the spinning D6-brane
theory, for which a stable region can be found, is presented in
detail.
Corresponding results for the spinning M5-M2 brane bound state,
including the near-horizon limit and thermodynamics, are given as
well.}

\begin{document}

\maketitle %%%%%%%%%% THIS IS IGNORED %%%%%%%%%%%
% \newpage

%---------------------------
\section{Introduction}

Non-commutative geometry appears naturally in certain limits of
string theory with a background NSNS $B$-field, as first
discovered in the context of M(atrix) theory \cite{Connes:1998cr}.
Recently, it has been shown \cite{Seiberg:1999vs} that
non-commutative super Yang-Mills (NCSYM) theory   directly appears
from open string interactions, as suggested in earlier studies
\cite{Douglas:1998fm} of the subject. 
More specifically, NCSYM appears \cite{Seiberg:1999vs} in a special
low-energy limit of the world-volume theory of $N$ coinciding
D$p$-branes in the presence of a NSNS $B$-field. 
This fact has been used to extend the
correspondence between near-horizon D$p$-brane supergravity
solutions and super Yang-Mills (SYM) theories in $p+1$ dimensions
\cite{Maldacena:1997re,Itzhaki:1998dd,Aharony:1999ti}, to a correspondence
between near-horizon D$p$-brane supergravity solutions with a
non-zero NSNS $B$-field and NCSYM in $p+1$ dimensions
\cite{Maldacena:1999mh,Hashimoto:1999ut,Li:1999am,Alishahiha:1999ci}.
See also Refs. \cite{Bigatti:1999iz,Barbon:1999mx,Cai:1999aw,Mikhailov:1999fd,Hashimoto:1999yj}
for further recent and related studies 
of non-commutative geometry in 
string theory.  

 The purpose of this
paper is two-fold:\newline (i) Extend the analysis of the NCSYM
phase structure given in Ref. \cite{Alishahiha:1999ci}, to a more
general path in the gauge theory phase space and use this to study
the validity of D-brane thermodynamics for the NCSYM. \newline
(ii) Construct spinning D-brane bound state solutions and use
their near-horizon limit to analyze the thermodynamics of NCSYM,
extending our recent work \cite{Harmark:1999xt}. Modification of
the gauge coupling phase structure of (i) due to the rotation will
be considered as well.
 Since angular momenta and
velocities on the supergravity side correspond to R-charges and
R-voltages on the NCSYM side, the thermodynamics of these spinning
brane solutions with a $B$-field may provide further insights into
NCSYM.

We start in Section \ref{secdbrane} with constructing general
spinning D-brane bound state solutions, by applying a set of
T-dualities to the general spinning D$p$-brane solutions
\cite{Cvetic:1999xp,Harmark:1999xt}. The resulting
backgrounds\footnote{For zero angular momentum these spinning
bound state solutions reduce to the bound states given in Refs.
\cite{Russo:1997if,Breckenridge:1997tt}.} are bound states of
spinning $D(p-2k)$-branes, $k=0 \ldots m$, with $2m$ the rank of
the NSNS $B$-field. For a given $p$, the solution is spinning in the $9-p$
dimensional transverse space, and we show that, except for
additional charges and chemical potentials of the lower branes in
the bound state, the thermodynamics is equivalent to that of a
spinning D$p$-brane.

In Section \ref{secnhlimit} we construct the near-horizon limit of
the general spinning D-brane bound state, and discuss the
conditions in order for the near-horizon solution to describe the
dual NCSYM, focusing for simplicity on the non-rotating case
first. The phase space of the NCSYM at fixed $N$ is parametrized
by the YM coupling constant $\gym$, the gauge theory energy scale
$r$ and the non-commutativity parameters $b_k$, $k=1 \ldots m$,
which enter the position commutators
\cite{Maldacena:1999mh,Alishahiha:1999ci}
\begin{equation}
[ y^{2k-1}, y^{2k}] =i  b_k \sp k = 1 \ldots m
\end{equation}
By considering a general type of path in phase space, we discuss
certain general features of the resulting phase diagrams in terms
of the effective coupling $\geff$. This analysis depends on the
dimension $p$ of the D$p$-brane and on the rank $2m$ of the NSNS
$B$-field. We find four types of phase diagrams, and analyze in
detail which phase diagram is relevant for the chosen path and
region of phase space. For each value of $p$ and $m>0$, we
establish  that a path and region of phase space can be chosen
such that the phase structure of any of the four phase diagrams
can be realized. We also determine under which conditions the
description with finite $N$ is valid.
Our analysis includes the case considered
in \cite{Alishahiha:1999ci} and, as another
case of special interest, the path in phase space along which
the intensive thermodynamic quantities are invariant,
which is used to examine the validity of the thermodynamics.
We end
Section \ref{secnhlimit} by discussing the effects of angular
momenta on the gauge coupling phase structure.

In Section \ref{secnhtherm} we discuss the thermodynamics of the
near-horizon solutions and their corresponding dual NCSYM
theories. As was argued for the non-rotating case
\cite{Maldacena:1999mh,Alishahiha:1999ci,Barbon:1999mx,Cai:1999aw},
 it is seen that the thermodynamic
quantities are the same as for commutative SYM case, with the
coupling constant $\gym^2$ of the latter replaced by $\gym^2
\prod_{k=1}^m b_k$ in the NCSYM case. Then, using the results of
Section \ref{secnhlimit}, the range of validity of the
thermodynamics is discussed,  explaining specifically the cases
for which a) the coupling can go all the way to infinity and b)
finite $N$ is allowed. Moreover, we find the region of
thermodynamic phase space where both these properties can hold. We
conclude this section with a detailed analysis of the
thermodynamics of the spinning D6-brane theory, adding to the
analysis of the near-horizon D6-brane  in Ref.
\cite{Harmark:1999xt}. This is of interest in view of the recent
discovery that the D6-brane theory decouples from gravity for $m
\geq 1$ \cite{Maldacena:1999mh,Alishahiha:1999ci}, and since there
is a critical angular momentum density above which the spinning
D6-brane  is stable in the canonical ensemble
\cite{Harmark:1999xt}.

For completeness, Section \ref{secm5}  gives the corresponding
results for the spinning M5-M2 brane bound state, which can be
obtained by lifting the spinning D4-D2 brane bound state to M-theory. The
asymptotically-flat solution is given with its thermodynamics as
well as the near-horizon solution, which is dual to the
non-commutative (2,0) theory. As expected the thermodynamics is
again independent of the non-commutativity parameter. Finally,
Section \ref{seccon} presents conclusions and discussion.
Appendix \ref{apptdual} reviews the T-duality transformations used
to obtain our spinning bound state solutions and Appendix
\ref{apprrpot} provides some details on the RR gauge potentials of
these solutions and their near-horizon limit.

%---------------
\section{Spinning D-brane bound states \label{secdbrane} }

The spinning D-brane bound states that we will present are
solutions of the low-energy effective action of type II string
theory in the string frame
\begin{equation}
\label{pbact}
 I = \frac{1}{16\pi G} \int d^{10} x \sqrt{g} \left[
e^{-2\phi} \Big(  R + 4 \partial_\mu \phi \partial^\mu \phi -
\frac{1}{12} H_3^2 \Big) - \sum_p \frac{1}{2 (p+2)!} F_{p+2}^2
\right] + I_{WZ}
\end{equation}
Here \( H_3 = dB_2 \) and \( F_{p+2} = dA_{p+1} + \ldots \) are
 the field strengths of the NSNS 2-form and $(p+1)$-form RR gauge
 potentials respectively, where the dots denote terms involving
 $B_2$. The integers $p$ are even (odd) for type IIA (IIB) and $I_{WZ}$
 denotes topological terms involving $B_2$ and $A_{p+1}$.
Using T-duality as a solution generating technique (see Appendix
\ref{apptdual}) one finds from the general spinning D-branes
\cite{Cvetic:1999xp,Harmark:1999xt} the corresponding spinning
D-brane bound states
 with a non-zero $B$-field. In the notation of
\cite{Harmark:1999xt} these new solutions take the following form:
The metric is
\begin{eqnarray}
\label{solmet}
 ds^2 &=& H^{-1/2} \Big( - f dt^2 +
\sum_{k=1}^m D_k \Big[ (dy^{2k-1})^2 + (dy^{2k})^2 \Big] +
\sum_{i=2m+1}^p (dy^i)^2  \Big) \nn \\ & & + H^{1/2} \Big(
\bar{f}^{-1} K_{9-p}\ dr^2 + \Lambda_{\alpha \beta} d\eta^\alpha
d\eta^\beta \Big) \nn
\\ && + H^{-1/2} \frac{1}{W_p} \frac{r_0^{7-p}}{r^{7-p}} \Big(
\sum_{i,j=1}^n l_i l_j \mu_i^2 \mu_j^2 d\phi_i d\phi_j - 2 \cosh
\alpha  \sum_{i=1}^n l_i \mu_i^2 dt d\phi_i \Big)
\end{eqnarray}
while the dilaton takes the form
\begin{equation}
\label{soldil} e^{2\phi} = H^{(3-p)/2} \prod_{k=1}^{m} D_k
\end{equation}
The NSNS $B$-field has rank $2m \leq p$ and is given by
\begin{equation}
\label{solb}
 B_{2k-1,2k} = \tan \theta_k \Big( H^{-1} D_k - 1
\Big) \sp k=1 \ldots m
\end{equation}
and we refer to Appendix \ref{apprrpot} for the form of the
non-zero RR gauge potentials $A_{p-2k+1}$, $k =0 \ldots m$.

The functions\footnote{In Ref. \cite{Harmark:1999xt} the functions
$K_d, L_d$ are  labeled by the transverse dimension $d$, which is
equal to $9-p$  for D$p$-branes, and we have kept the same
definitions for these.}
entering this background are given by
\begin{subequations}
\label{defs}
\begin{equation}
\label{lhdef}
 L_{9-p} = \prod_{i=1}^n \Big( 1 + \frac{l_i^2}{r^2}
\Big) \sp H = 1 + \frac{1}{W_p} \frac{r_0^{7-p} \sinh^2
\alpha }{r^{7-p}}
\end{equation}
\begin{equation}
f = 1 - \frac{1}{W_p} \frac{r_0^{7-p}}{r^{7-p}} \sp \bar{f} = 1 -
\frac{1}{L_{9-p}} \frac{r_0^{7-p}}{r^{7-p}}
\end{equation}
\end{subequations}
where
\begin{equation}
\label{Np}
 W_p = K_{9-p} L_{9-p}
\end{equation}
with $L_{9-p}$ defined in \eqref{lhdef} and $K_{9-p}$ entering the
flat transverse space metric
\begin{equation}
\label{flattr} \sum_{a=1}^{9-p} (d x^a)^2 = K_{9-p}\ dr^2 +
\Lambda_{\alpha \beta} d\eta^\alpha d\eta^\beta
\end{equation}
(See Appendix B of  Ref. \cite{Harmark:1999xt} for the explicit
expressions of $K_{p-9}$, $\Lambda_{\alpha \beta}$ and the $\mu_i$
in \eqref{solmet}). We have also defined
\begin{equation}
\label{ddef} D_k = \Big( \sin^2 \theta_k H^{-1} + \cos^2 \theta_k
\Big)^{-1} \sp k=1 \ldots m
\end{equation}
Finally we recall the definition
\begin{equation}
\label{hdef} h^{7-p} = r_0^{7-p} \cosh \alpha \sinh \alpha 
\end{equation}
and the relations
\begin{equation}
\label{IIrel}
16 \pi G = (2\pi)^7 g_s^2 l_s^8
\sp
h^{7-p} = \frac{(2\pi)^{7-p} N g_s l_s^{7-p}}{(7-p) V(S^{8-p})}
\prod_{k=1}^m (\cos \theta_k)^{-1}
\end{equation}
where $l_s$ is the string length, $g_s$ the string coupling and
 $V(S^{8-p})$ the volume of the unit $(8-p)$-sphere. The second
 relation in \eqref{IIrel} is a consequence of charge
 quantization of the D$p$-brane, where $N$ is the number of coincident
D$p$-branes.

These solutions represent spinning bound states of
those branes
that carry charges under the non-zero RR fields, i.e.
of D$(p-2k)$, $k=0 \ldots m$. The background depends
 on the
non-extremality parameter $r_0$, charge parameter $\alpha $, the
angular momenta $l_i$, $i = 1 \ldots n$ ($n \equiv [(9-p)/2]$) and
the angles $\theta_k$, $k=1 \ldots m$. For zero angular momentum
the bound state  solutions of
\cite{Russo:1997if,Breckenridge:1997tt} are recovered.

Besides the charges and chemical potentials, the thermodynamic quantities of
the bound state solution are not affected by
the non-zero $B$-field in that they are given by the corresponding
expressions of the spinning D$p$-brane
\cite{Harmark:1999xt}
\begin{subequations}
\label{pbranethermo}
\begin{equation}
\label{admmass}
M = \frac{V_p V(S^{8-p})}{16\pi G} r_0^{7-p}
\Big( 8-p + (7-p)\sinh^2 \alpha \Big)
\end{equation}
\begin{equation}
\label{tands}
T = \frac{7-p-2\kappa }{4 \pi r_H \cosh \alpha } \sp
S = \frac{V_p V(S^{8-p}) }{4G} r_0^{7-p} r_H \cosh \alpha 
\end{equation}
\begin{equation}
\label{angmom}
\Omega_i = \frac{l_i}{(l_i^2 + r_H^2)\cosh \alpha } \sp
J_i = \frac{V_p V(S^{8-p})}{8\pi G} r_0^{7-p} l_i \cosh \alpha 
\end{equation}
\end{subequations}
We refer to Appendix \ref{apprrpot} for the expressions of
the charges and chemical potentials, which satisfy
\begin{subequations} \label{summuQ}
\begin{equation}
  \sum_{k=0}^{m} \sum_\alpha
   \mu_{p-2k}^{(\alpha)} Q_{p-2k }^{(\alpha)}= \mu Q
\end{equation}
\begin{equation}
\label{muandq0} \mu = \tanh \alpha \sp Q = \frac{V_p
V(S^{8-p})}{16\pi G} r_0^{7-p} (7-p) \sinh \alpha \cosh \alpha 
\end{equation}
\end{subequations}
where for a given $k$, $\alpha$ labels the
distinct D$(p-2k)$ branes embedded in the D$p$-brane.
 In \eqref{pbranethermo} the horizon
radius $r_H$ and the coefficient $\kappa$ are given by
\begin{equation}\label{horkappa}
\prod_{i=1}^n \left( 1 + \frac{l_i^2}{r_H^2}\right)  r_H^{7-p} =
r_0^{7-p} \sp
 \kappa = \sum_{i=1}^n \frac{l_i^2}{l_i^2 + r_H^2 }
\end{equation}
The first law of thermodynamics reads
\begin{equation}
\label{firstlaw}
 d M = T dS + \sum_{i=1}^n \Omega_i dJ_i +
\sum_{k=0}^{m} \sum_\alpha \mu_{p-2k}^{(\alpha)} d
Q_{p-2k}^{(\alpha)}
 \sp M=M(S,\{ J_i \},\{Q_{p-2k}^{(\alpha)} \} )
\end{equation}
 It then follows from
\eqref{pbranethermo} and \eqref{summuQ} that the integrated Smarr formula
\begin{equation}
(7-p) M = (8-p) TS + (7-p) \mu  Q + (8-p) \sum_{i=1}^n \Omega_i
J_i
\end{equation}
is satisfied.

Note that from the D3-D1 or D5-D3-D1 bound state, obtained for
$p=3$ or 5, it is not difficult to obtain the corresponding
background of D3-NS1 or NS5-D3-NS1 using type IIB S-duality. The
thermodynamic quantities will then remain unchanged. For zero
angular momentum these solutions can be found in \cite{Lu:1999uv}
and \cite{Alishahiha:1999ci}.

%-------------------------------
\section{The near-horizon limit \label{secnhlimit} }

In this section we consider the near-horizon limit of the spinning
D-brane bound state solutions found in Section \ref{secdbrane}. In
Section \ref{secnearhorsol} we obtain the near-horizon solution by
taking the appropriate limit. In Section \ref{secdualtheor} we
review some properties of the dual non-commutative
field theories corresponding to
these near-horizon solutions, while  Section \ref{secphasestruc}
gives a detailed analysis of the gauge coupling phase structure.
Finally, Section \ref{secangmom} describes the effect of non-zero
angular momenta in the dual field theory, and discusses the
induced modifications in the analysis of the gauge coupling phase
structure, presented in Section \ref{secphasestruc} for zero
angular momenta.

%-------------------------------------
\subsection{Near-horizon solutions
\label{secnearhorsol} }

We start by  constructing the near-horizon limit of the spinning
bound state \eqref{solmet}-\eqref{solb} in which  the magnetic
field \eqref{solb} is taken to infinity in such a manner that a
finite rescaled value is obtained after taking the limit. As
reviewed in Section \ref{secdualtheor}, this corresponds to a
non-commutative field theory on the world-volume of the
D$p$-brane. For the non-rotating case, this limit was found in
Refs.
\cite{Seiberg:1999vs,Maldacena:1999mh,Hashimoto:1999ut,Alishahiha:1999ci}.
 This limit crucially
depends on the rank of the $B$-field, denoted by $2m \leq p $.

The near-horizon limit is defined by letting the string length
 \( l_s \rightarrow 0 \) accompanied by the rescalings
\begin{subequations}
\label{rescal1}
\begin{eqnarray}
&&
r  = \frac{\old{r}}{l_s^2}   \sp
r_0  = \frac{\oldp{r_0}}{l_s^2}  \sp
l_i = \frac{\oldp{l_i}}{l_s^2} \sp
h^{7-p} = \frac{\old{h}^{7-p}}{l_s^{10-2p}}
\\ &&
\label{fieldscal}
 ds^2 = \frac{\oldp{ds^2}}{l_s^2} \sp e^{\phi} =
l_s^{3-p+2m} e^{\phi_{\rm old}} \sp G  =
\frac{\old{G}}{l_s^{14-2p+4m}}
\end{eqnarray}
\end{subequations}
and the rescalings
\begin{subequations}
\label{rescal2}
\begin{eqnarray}
&&
b_k = l_s^2 \tan \theta_k   \sp
y^{2k-1} = \frac{b_k}{l_s^2} \oldp{y^{2k-1}}    \sp
y^{2k} = \frac{b_k}{l_s^2} \oldp{y^{2k}}
\\ &&
\label{bscal}
 B_{2k-1,2k} = l_s^{-2} \oldp{B_{2k-1,2k}} \sp k=1
\ldots m
\end{eqnarray}
\end{subequations}
where the quantities on the left-hand side in \eqref{rescal1},
\eqref{rescal2} are kept fixed, and the quantities on the
right-hand side (labelled with subscript ``old'' except for
$\theta_k$) are those that enter the asymptotically-flat solutions
of Section \ref{secdbrane}. The corresponding rescalings of the RR
gauge potentials are given in \eqref{rrrescal}, which  together
with \eqref{fieldscal}, \eqref{bscal} leave the low-energy
effective action \eqref{pbact} invariant. Note that for $m=0$
(i.e. zero $B$-field) the rescalings \eqref{rescal1} correctly
reduce to the rescalings of the spinning D$p$-brane solutions
described in \cite{Harmark:1999xt}.

Taking the near-horizon limit, the following near-horizon spinning
D$p$-brane solution with non-zero $B$-field is obtained
\begin{subequations}
\begin{eqnarray} \label{nhmet} && ds^2 = H^{-1/2} \left( - f
dt^2 + \sum_{k=1}^m D_k \Big[ (dy^{2k-1})^2 + (dy^{2k})^2 \Big] +
\sum_{i=2m+1}^p (dy^i)^2 \right) \nn \\ && + H^{1/2} \Big(
\bar{f}^{-1} K_{9-p} dr^2 + \Lambda_{\alpha \beta} d\eta^\alpha
d\eta^\beta \Big)
- 2 H^{-1/2} \frac{h^{\frac{7-p}{2}}r_0^{\frac{7-p}{2}}}{W_p r^{7-p}}
\sum_{i=1}^n l_i \mu_i^2 dt d\phi_i
\end{eqnarray}
\begin{equation}
\label{nhdil}
e^{2\phi} = H^{(3-p)/2} \prod_{k=1}^{m} b_k^2 D_k
\end{equation}
\begin{equation}
B_{2k-1,2k} = \frac{1}{b_k} \frac{a_k^{7-p} W_p
r^{7-p}}{1+a_k^{7-p} W_p r^{7-p}} \sp k=1 \ldots m
\end{equation}
\end{subequations}
where now
\begin{equation}
H = \frac{h^{7-p}}{W_p r^{7-p}}   \sp D_k = \Big( 1 + a_k^{7-p}
W_p r^{7-p} \Big)^{-1} \sp k=1 \ldots m
\end{equation}
and we have defined
\begin{equation}
\label{adef}
 a_k^{7-p} = \frac{b_k^2}{h^{7-p}}\sp k=1 \ldots m
\end{equation}
The functions $K_{9-p}$, $W_p$ and $\Lambda_{\alpha \beta}$ are
not affected by the rescaling and hence as in Section
\ref{secdbrane}. Note that a gauge transformation has been made
that removes the constant part in $B$.
 The corresponding expression for the RR gauge
potentials of the near-horizon solution can be obtained in
principle with the data of Appendix \ref{apprrpot}. For zero
angular momenta the background reduces to the near-horizon
solutions of
\cite{Maldacena:1999mh,Hashimoto:1999ut,Alishahiha:1999ci}.

%----------------------------------------
\subsection{The dual field theories
\label{secdualtheor} }

We continue with describing the map between the variables of the
near-horizon supergravity solution and the
dual field theory variables. This is done for zero angular momenta ($l_i =0$),
but in Section \ref{secangmom} the modifications arising
from non-zero \( l_i \) will be discussed.
We also comment on the validity of the thermodynamics of the dual
field theories and review the conditions under which gravity
decouples from the world-volume theory on the brane  in the
near-horizon limit.

 The zero slope limit \( l_s \rightarrow 0 \)
in the presence of a $B$-field with rank $2m > 0$, considered in
Section \ref{secnearhorsol}, gives at low energies a world-volume
theory that is described by the D$p$-brane Born-Infeld action with
a non-zero $B$-field. The latter  is equivalent
\cite{Seiberg:1999vs} to a non-commutative supersymmetric
Yang-Mills (NCSYM) theory in $p+1$ dimensions  with 16
supercharges, where the non-commutativity of the coordinate pair
$(y^{2k-1},y^{2k})$ is given by \( [ y^{2k-1}, y^{2k} ] =i  b_k
\) \cite{Seiberg:1999vs,Maldacena:1999mh,Alishahiha:1999ci}. The
NCSYM theory on the brane has a coupling constant $\gym$ given by
\begin{equation}
\label{ncgym}
 \gym^2 = (2\pi)^{p-2} g_s l_s^{p-3-2m} =
(2\pi)^{p-2} \bar{g}_s \sp \bar{g}_s = g_s l_s^{p-3-2m}
\end{equation}
where \( \bar{g}_s \) is the rescaled string coupling constant.
{}From \eqref{IIrel} and the rescalings
\eqref{rescal1}, \eqref{rescal2} we find in the near-horizon limit
the new quantities
\begin{equation}
\label{IIrelnh}
h^{7-p} = \frac{(2\pi)^{9-2p} \gym^2 N
\prod_{k=1}^m b_k}{(7-p) V(S^{8-p})} \sp 16 \pi G = (2\pi)^7
\bar{g}_s^2 = (2\pi)^{11-2p} \gym^4
\end{equation}
Following \cite{Itzhaki:1998dd,Alishahiha:1999ci} we introduce
the effective gauge coupling of the world-volume theory
\begin{equation}
\label{geff}
 \geff^2 = \gym^2 N \left( \prod_{k=1}^m b_k \right)
r^{p-3}
\end{equation}
where the rescaled radius \( r \) has the interpretation of the
effective energy scale of the field theory, being the expectation
value of the Higgs field \cite{Maldacena:1997re,Itzhaki:1998dd}.
Since the curvature of the metric \eqref{nhmet} is of order \( 1 /
\geff \), the requirement that  curvatures be small imposes the
restriction  \( \geff \gg 1 \), so that one needs to be in the
strong coupling region for the supergravity description of the
D-brane world-volume theory to hold. The perturbative description
of the world-volume theory on the D-brane is instead valid at weak
coupling  \( \geff \ll 1 \).

We also introduce the effective non-commutativity (NC) parameters
\cite{Alishahiha:1999ci}
\begin{equation}
\label{aeff}
 \aeff_k = a_k r \sp k=1 \ldots m
\end{equation}
Then, the region \( \aeff_k \ll 1 \)  corresponds to the
coordinate pair \( (y^{2k-1}, y^{2k}) \) being commutative, while
in the opposite region \( \aeff_k \gg 1 \) these coordinates are
non-commutative. We finally introduce the effective string
coupling \cite{Itzhaki:1998dd,Alishahiha:1999ci}
\begin{equation}
\label{gseff}
 \gseff = \bar{g}_s e^\phi
\end{equation}
and from \eqref{nhdil} and \eqref{ncgym} it follows that
\begin{equation}
\label{gsregions}
\gseff \ll 1 \ \Leftrightarrow \
\geff^2 \ll N^{\frac{4}{7-p}} \prod_{k=1}^m
\Big( 1 + (\aeff_k)^{7-p} \Big)^{\frac{2}{7-p}}
\end{equation}
The fact that the NC parameters \( \aeff_k \) enter in the
effective string coupling \cite{Alishahiha:1999ci} has interesting
consequences, as we will see shortly.

The requirements $\geff \gg 1$ and \eqref{gsregions} in principle
determine for which values of $N$, $\gym$, $r$ and $b_k$, $k=1
\ldots m$, the dual field theory is described by the D-brane in
the near-horizon limit. For the sequel, it is important to note
that the D-brane thermodynamics in the near-horizon limit gives
the thermodynamics of the dual field theory precisely when we set
$r=r_0$ in the relevant expressions \cite{Itzhaki:1998dd}. In the
special case of the D3 and D5-brane, we note that the condition \(
\gseff \ll 1 \) (with $r=r_0$) is no longer relevant for the
thermodynamics, since their S-dual branes, being the D3 and
NS5-brane, have the same thermodynamics\footnote{This is explained
for the spinning D5 and NS5-brane in Ref. \cite{Harmark:1999xt}.}.
As a consequence, for the D3 and D5-brane case  there is no bound
on \( \gseff \) as far as the thermodynamics is concerned.

It is also important to note that for the D$p$-branes with $p \leq
5$ gravity decouples from the D-brane world-volume theory for all
values of $m$, while for the D6-brane  
 this only happens when $m
\geq 1$ \cite{Maldacena:1999mh,Alishahiha:1999ci}. In other words,
for the D6-brane the presence of a non-zero $B$-field is crucial
for the decoupling of gravity. Only when gravity decouples can we
expect to have a well-defined field theory that is described by
supergravity when the curvatures are small. As noted in
\cite{Maldacena:1999mh,Alishahiha:1999ci} the decoupling of
gravity for the D6-brane with $m \geq 1$ implies the existence of
a 7 dimensional non-commutative field theory. For this reason we
will, after having discussed the thermodynamics and phase
structure in generality, give a more detailed  account in Section
\ref{secd6} of what we can infer for this D6-brane theory.

%--------------------------
\subsection{Gauge coupling phase structure
\label{secphasestruc} }

We now proceed to find the gauge coupling phase structure of the
D-brane world-volume theories in a more general setting than
previously done. In \cite{Itzhaki:1998dd} the gauge coupling phase
structure was considered for the D-brane world-volume theories
without $B$-field, that is, with $m=0$. As we shall see, for $m>0$ a
much richer structure is observed. The phase diagrams obtained
in \cite{Alishahiha:1999ci} for $m > 0$ for the D2, D4, D5 and
D6-brane will arise as a special case of the analysis we present
here. In the following we take \( 2 \leq p \leq 6 \) and \( 1 \leq
m \leq [\frac{p+1}{2}] \)\footnote{Note that the analysis of this
section also holds for \( m = \frac{p+1}{2} \) if $p$ is odd, when
the Euclidean background is considered as done in
\cite{Alishahiha:1999ci}.}. We also set the angular momenta to
zero, commenting in Section \ref{secangmom} on the modifications
due to non-zero \( l_i \).

For simplicity, we set $b_k = b$, $k=1 \ldots m$, so that
  the phase space is  parameterized by $N$,
$\gym$, $r$ and $b$. Our aim is to study the phase structure in
terms of the effective gauge coupling
\begin{equation}
\label{xdef}
\zz \equiv \geff^2
\end{equation}
going from zero to infinity. However, there is actually
a large freedom in the choice of path in phase space that one
may follow, which we parametrize by
\begin{equation}
\label{phasepath} \gym^2 \propto z^\alpha \sp r \propto z^\beta
\sp b \propto z^\gamma
\end{equation}
keeping $N$ fixed. From  \eqref{geff} and the definition
\eqref{xdef}  it then follows that the scaling exponents $\alpha$,
$\beta$, $\gamma$ obey the constraint
\begin{equation}
\label{abc}
 \alpha + (p-3) \beta + m \gamma = 1
\end{equation}
It also follows from \eqref{adef}, \eqref{IIrelnh}-\eqref{aeff} that as a function of $\zz$ we can write
\begin{equation}
\label{aeffx}
(\aeff)^{7-p} = \left( \frac{\zz}{\xnc} \right)^\eta
\end{equation}
where
\begin{equation}
\label{eta}
\eta = 4\beta + 2\gamma -1
\end{equation}
and $\xnc$ is a constant determined by \eqref{aeffx}. Since
$\aeff$ determines the (non)-commutativity of the theory, we find
that for $\eta> 0$  the field theory is non-commutative for $\zz \gg
\xnc$ and commutative for $\zz \ll \xnc$, while in the case
 $\eta < 0$ the field theory is commutative for $\zz \gg \xnc$
and non-commutative for $\zz \ll \xnc$.

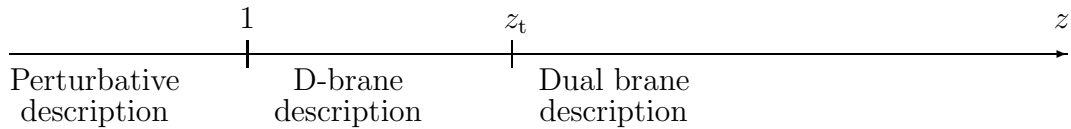
\begin{figure}[h]
\begin{picture}(410,60)(0,0)
\put(10,35){\vector(1,0){400}}
\put(405,45){$\zz$}
\put(10,10){\shortstack{Perturbative \\ description}}
\put(110,10){\shortstack{D-brane \\ description}}
\put(210,10){\shortstack{Dual brane \\ description}}
\put(97,45){1}
\put(100,30){\line(0,1){10}}
\put(197,45){$\xt$}
\put(200,30){\line(0,1){10}}
\end{picture}
\caption{Phase diagram with one transition point \( \xt \)
where \( \gseff = 1 \). \label{figonext} }
\end{figure}

{}Using \eqref{aeffx} the condition \eqref{gsregions} can be
rewritten as
\begin{equation}
\label{gsx}
\gseff \ll 1 \ \Leftrightarrow
\ \zz  \ll N^{\frac{4}{7-p}}
\left( 1 + \Big( \frac{\zz}{\xnc} \Big)^\eta \right)^{\frac{2m}{7-p}}
\end{equation}
which will enable us to  analyze the phase structure in terms of
$\zz$. One of the common features of this phase structure is that
for $\zz \ll 1$ we have a perturbative description of the
world-volume field theory, while for $\zz \gg 1$ the curvatures
are small and the supergravity description of the world-volume
theory is valid.  If there is a transition point $\xt$ where \(
\gseff = 1 \) then the supergravity D$p$-brane description is
valid in the range \( 1 \ll \zz \ll \xt \) and for \( \zz \gg \xt
\) we go to an S-dual brane description. In type IIB string theory
this means the NS5-brane for the D5-brane, or the D3-brane itself
in the case of the D3-brane. For type IIA string theory this means
a supergravity solution up-lifted to 11-dimensional supergravity
\footnote{For the D2-brane we have an additional phase transition
point in the 11-dimensional sector, namely the point where the
supergravity solution becomes a localized M2-brane in 11
dimensions \cite{Itzhaki:1998dd}.}. As described in
\cite{Itzhaki:1998dd} the world-volume field theories on the
D-branes with $m=0$ all have the phase structure with one
transition point \( \xt \sim N^{\frac{4}{7-p}} \) where \( \gseff
= 1 \). The phase diagram corresponding to this phase structure is
depicted in Figure \ref{figonext}.

We divide the description of the gauge coupling phase structure
into five cases, depending on the parameter $\eta$. Except for
$\eta=0$, each case is again subdivided into the two cases $\xnc
\ll N^{\frac{4}{7-p}}$ and $\xnc \gg  N^{\frac{4}{7-p}}$:

\begin{enumerate}
\item $ \eta < 0 $ : Here $\zz \ll \xnc$ is the non-commutative sector,
so that $\zz \ll 1$ gives  a perturbative NCSYM description
(assuming that $\xnc > 1$).

Consider the case \( \xnc \ll N^{\frac{4}{7-p}} \).
This means that the point \( \zz = N^{\frac{4}{7-p}} \) lies in the
commutative sector, so the supergravity D$p$-brane description
is valid for \( 1 \ll \zz \ll N^{\frac{4}{7-p}} \), and hence we need
\( N \gg 1 \). For \( \zz \gg N^{\frac{4}{7-p}} \) we go to the
S-dual brane theory and since we are in the commutative sector
the phase structure is the same as found in \cite{Itzhaki:1998dd}
and the phase diagram is depicted in
Figure \ref{figonext} with \( \xt \sim N^{\frac{4}{7-p}} \).

Consider the case \( N^{\frac{4}{7-p}} \ll \xnc \).
Here the transition point $\xt$ where $\gseff = 1$ is
between \( N^{\frac{4}{7-p}} \) and \( \xnc \).
Except for the shift in \( \xt \) the phase structure
is the same as for \( \xnc \ll N^{\frac{4}{7-p}} \), thus
the phase diagram is again Figure \ref{figonext}.
Remarkably, we can have \(N\) finite in this case.

\item   $\eta = 0$ : In this case \( \aeff \) in \eqref{aeffx} is constant.
There is  a transition point \( \xt \sim N^{\frac{4}{7-p}}
(1+(\aeff)^{7-p})^{\frac{2m}{7-p}} \) for which \( \gseff = 1 \).
Apart from a shift in \( \xt \) the phase structure is
qualitatively the same as for $m=0$, and thus as depicted in Figure
\ref{figonext}. This case is interesting since we can choose \(
\aeff \gg 1 \) so that the theory is non-commutative for all
$\zz$. We can clearly have $N$ finite if we choose $\aeff \gg 1$.

\item   $0 < \eta < \frac{7-p}{2m}$ : Here $\zz \gg \xnc$ is the
non-commutative sector, so that $\zz \ll 1$ gives  a perturbative
commutative SYM description (assuming that $\xnc > 1$).

Consider the case \( N^{\frac{4}{7-p}} \ll \xnc \).
For this case the point \( N^{\frac{4}{7-p}} \) lies in the commutative
sector so the supergravity D$p$-brane description is valid
for \( 1 \ll \zz \ll N^{\frac{4}{7-p}} \), so that we need to require
\( N \gg 1 \).
Thus, apart from the non-commutativity the phase structure is the
same as for $m=0$, depicted in Figure \ref{figonext} with
\( \xt \sim N^{\frac{4}{7-p}} \).

Consider the case \( \xnc \ll N^{\frac{4}{7-p}} \). Here a
 transition point \( \xt \) is found where \( \gseff = 1 \),
and $\xt$ is either of the same order or larger than
$N^{\frac{4}{7-p}}$. Again the phase structure is qualitatively
the same as for $m=0$, depicted in Figure \ref{figonext}. If \(
\xnc \ll 1 \) it is possible to have $N$ finite (for small $\eta$
it should be $\xnc^{\eta} \ll 1$).

\item   $\eta = \frac{7-p}{2m}$ : Here $\zz \gg \xnc$ is the
non-commutative sector,
so for $\zz \ll 1$ we have a perturbative commutative
SYM description (assuming that $\xnc > 1$).

Consider the case \( N^{\frac{4}{7-p}} \ll \xnc \).
The supergravity D$p$-brane description is valid for
\( 1 \ll \zz \ll N^{\frac{4}{7-p}} \) and the string coupling will
become constant for large $\zz$. Except for the non-commutativity,
the phase structure is the same as for $m=0$,
depicted in Figure \ref{figonext}, with \( \xt \sim N^{\frac{4}{7-p}} \).
We need \( N \gg 1 \).

Consider the case \( \xnc \ll N^{\frac{4}{7-p}} \).
In this case we have that \( \gseff \ll 1 \) for all \( \zz \gg 1 \).
Thus, the supergravity D$p$-brane description is valid for
all \( \zz \gg 1 \). The phase diagram for this case in depicted
in Figure \ref{fignoxt}.
If \( \xnc \ll 1 \) it is possible to have $N$
finite.

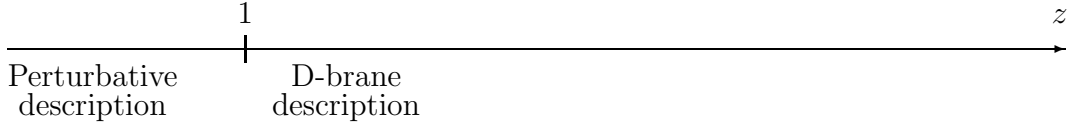
\begin{figure}[h]
\begin{picture}(410,60)(0,0)
\put(10,35){\vector(1,0){400}}
\put(405,45){$\zz$}
\put(10,10){\shortstack{Perturbative \\ description}}
\put(110,10){\shortstack{D-brane \\ description}}
\put(97,45){1}
\put(100,30){\line(0,1){10}}
\end{picture}
\caption{Phase diagram for the case where the D-brane
description is valid for all \(\zz\gg 1\). \label{fignoxt} }
\end{figure}

\item   $\eta > \frac{7-p}{2m}$ : Here $\zz \gg \xnc$ is the
non-commutative sector,
so for $\zz \ll 1$ we have a perturbative commutative
SYM description (assuming that $\xnc > 1$).

Consider the case \( N^{\frac{4}{7-p}} \ll \xnc \).
If  \( N \gg 1 \), then it can
be seen from \eqref{gsx} that the
supergravity D$p$-brane description is not only valid
 for \( 1 \ll \zz \ll N^{\frac{4}{7-p}} \)
but also for \( \zz \gg \xt' \) where \( \xt' \gg \xnc \).
This interesting phase structure is depicted in the
phase diagram in Figure \ref{figtwoxt} with \( \xt \sim N^{\frac{4}{7-p}} \).

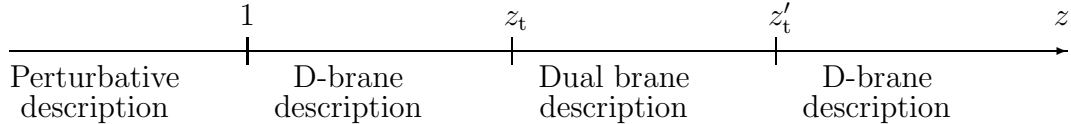
\begin{figure}[h]
\begin{picture}(410,60)(0,0)
\put(10,35){\vector(1,0){400}}
\put(405,45){$\zz$}
\put(10,10){\shortstack{Perturbative \\ description}}
\put(110,10){\shortstack{D-brane \\ description}}
\put(210,10){\shortstack{Dual brane \\ description}}
\put(310,10){\shortstack{D-brane \\ description}}
\put(97,45){1}
\put(100,30){\line(0,1){10}}
\put(197,45){$\xt$}
\put(200,30){\line(0,1){10}}
\put(297,45){$\xt'$}
\put(300,30){\line(0,1){10}}
\end{picture}
\caption{Phase diagram with two transition points \( \xt \)
and \( \xt' \) where \( \gseff = 1 \). \label{figtwoxt} }
\end{figure}

In the case of finite $N$, the transition at \( \zz=1 \)
goes from a perturbative field theory description into a supergravity
description with a brane configuration that is S-dual to the D$p$-brane.
At \( \xt \gg \xnc \) we have a transition into a D$p$-brane
description. The corresponding phase diagram is depicted
in Figure \ref{figonextdual}.

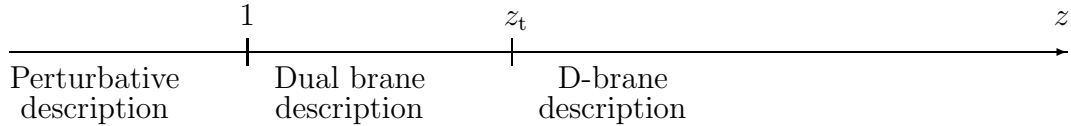
\begin{figure}[h]
\begin{picture}(410,60)(0,0)
\put(10,35){\vector(1,0){400}}
\put(405,45){$\zz$}
\put(10,10){\shortstack{Perturbative \\ description}}
\put(110,10){\shortstack{Dual brane \\ description}}
\put(210,10){\shortstack{D-brane \\ description}}
\put(97,45){1}
\put(100,30){\line(0,1){10}}
\put(197,45){$\xt$}
\put(200,30){\line(0,1){10}}
\end{picture}
\caption{Phase diagram with one transition point \( \xt \) where
\( \gseff = 1 \) but with large \( \gseff \) for \( \zz \ll \xt \)
and small \( \gseff \) for \( \zz \gg \xt \). \label{figonextdual}
}
\end{figure}

Consider the case \( \xnc \ll N^{\frac{4}{7-p}} \).
In this case we have that \( \gseff \ll 1 \) for all \( \zz \gg 1 \)
so that the D$p$-brane description is valid for all \(\zz \gg 1 \).
The phase diagram corresponding to this case is depicted in Figure
\ref{fignoxt}.
If \( \xnc \ll 1 \) it is possible to have $N$ finite.

\end{enumerate}

From the above analysis we infer the following general conclusions: \newline
(i) Comparing the analysis
to the phase structure of the $m=0$ case \cite{Itzhaki:1998dd}
a much richer structure is observed for $m>0$:
For $m=0$ there was basically only one type of
phase structure, namely the one depicted in Figure \ref{figonext}.
Instead, for $m>0$ there  are four types of phase diagrams,
depicted in Figures \ref{figonext}-\ref{figonextdual}. \newline
(ii) For each of the five cases, covering all values of $\eta$,
it is possible to find a regime with finite $N$. In fact, this
 regime corresponds to having the non-commutativity
as significant as possible, i.e. having the largest possible part
of the $\zz=\geff^2$ phase space non-commutative. Thus, it is
possible to have a dual supergravity description of NCSYM at
strong coupling for finite $N$. It is interesting to note that the 
scaling factor $N^2$ in the extensive thermodynamic quantities 
(see Section \ref{secthermquan}) persists for finite $N$.
This is different from the strong coupling SYM at finite $N$, which
has an $N^2-1$ factor instead, coming from the $SU(N)$ group.  
In Section \ref{seccon} we comment on the connection between these two
different factors.
\newline
(iii) The phase structure depicted in Figure \ref{fignoxt}
is particularly interesting since only one phase is present for \( \zz \gg 1\):
The string coupling constant is small for all \( \zz \gg 1\) so that
the D-brane description is valid in this entire range.
As one can extract from case 4 and 5 above, this phase structure occurs
when $\eta \geq \frac{7-p}{2m}$ and $\xnc \ll N^{\frac{4}{7-p}}$.
If in addition $\xnc \ll 1$ we can also have finite $N$.

We now consider some special choices of $\alpha$, $\beta$,
$\gamma$ in \eqref{phasepath}, using Eqs. \eqref{abc} and
\eqref{eta}. We are especially interested to find the cases that
allow $\eta \geq \frac{7-p}{2m}$ since this gives rise to a very
interesting phase structure.

\begin{enumerate}

\item[a)]  $\alpha = \gamma = 0$: This case corresponds to the one
described in Ref. \cite{Alishahiha:1999ci}, where the energy $r$ is
varied while keeping  all other parameters fixed. Clearly, we
cannot have $p = 3$, and for the other branes we have $\beta =
\frac{1}{p-3}$ and $\eta = \frac{7-p}{p-3}$. This means that $\eta
\geq \frac{7-p}{2m}$ is equivalent to $p > 3$ and $p-3 \leq 2m$.
This is fulfilled for $p=4,5$ for $m \geq 1$ and $p=6$ for $m\geq
2$, as was found in \cite{Alishahiha:1999ci}.

\item[b)]  $\beta = \gamma = 0$: In this case only the
YM coupling $\gym$ is varied. We have $\alpha = 1$ and $\eta =
-1$, so that for all cases the phase structure is the one depicted
in Figure \ref{figonext}.

\item[c)]  $\alpha = \beta = 0$: In this case we vary only the
NC parameter $b$. We have $\gamma = \frac{1}{m}$ and $\eta = \frac{2-m}{m}$
so we have $\eta \geq \frac{7-p}{2m}$ only for $m=1$ and $p=5,6$.

\item[d)]  $\alpha + m \gamma = (5-p) \beta$: With this choice the quantity
$r^{5-p} / h^{7-p}$ is fixed which is necessary to keep the
temperature $T$ fixed, as we shall see in Section
\ref{secrangeval}. It follows that  $\beta = \frac{1}{2}$ and
$\eta = 1 + 2\gamma$. Choosing $\gamma = 0$ (in order not to
change the position commutators of the non-commutative field
theory) we find that $\eta \geq \frac{7-p}{2m}$ for $p=3,4$ with
$m = 2$ and $p=5,6$ with $m \geq 1$. The importance of this case
will be clear in Section \ref{secrangeval} where it will be
considered in relation with the thermodynamics.

\end{enumerate}

We can also reverse the logic and ask whether it is possible for a
given $p$ and $m$ to find $\alpha$, $\beta$ and $\gamma$ such that
a specific value of $\eta$ is obtained. This is trivially seen to
be true, in fact, since for a given $\eta$ there are only two
restrictions \eqref{abc}, \eqref{eta} on the three scaling
exponents $\alpha$, $\beta$ and $\gamma$, leaving  a freedom of
choice in these exponents. For example, with the additional
constraint  $\gamma=0$, we find that $\alpha =
1-(p-3)\frac{1+\eta}{4}$ and $\beta = \frac{1+\eta}{4}$. It is
interesting to apply this to the D2-brane, since in
\cite{Alishahiha:1999ci} no deviation from the usual phase
structure was found in this case. For $p=2$ and $m=1$ we have
$\alpha = \frac{5+\eta}{4}$ and $\beta = \frac{1+\eta}{4}$ with
the choice $\gamma=0$. Thus, also for the D2-brane, all four types
of phase structures are possible.

%----------------------------------------------------
\subsection{Non-zero angular momenta \label{secangmom} }

In this section we extend the analysis of Sections
\ref{secdualtheor} and \ref{secphasestruc} to non-zero angular
momenta \( l_i \), $i=1...n$.

The isometry group $SO(9-p)$ of the transverse sphere of a
D$p$-brane corresponds to the R-symmetry group of the dual field
theory. From the point of view of  the thermodynamics, the Cartan
subgroup $SO(2)^n$ of the $SO(d)$ manifests itself as the
thermodynamic quantities $\{ \Omega_i \}$ corresponding to the
angular velocities in supergravity and the R-voltages in the dual
field theory, and $\{ J_i \}$  which are the angular momenta in
supergravity and the R-charges in the dual field theory (see e.g.
\cite{Gubser:1998jb,Kraus:1998hv,Cai:1998ji,Cvetic:1999rb,Harmark:1999xt}). 
The thermodynamics with
non-zero \( l_i \) will be considered in Section \ref{secnhtherm}.

The effective field theory parameters $\geff$ and $\aeff_k$,
$k=1\ldots m$, are as before in \eqref{geff}, \eqref{aeff} since
$r$ is still the Higgs expectation value of the brane probe.
However, the effective dilaton is now(up to a constant)
\begin{equation}
\label{effdilangmom}
\gseff \sim \frac{\geff^{\frac{7-p}{4}} W_p^{\frac{p-3}{4}}}{N \prod_{k=1}^m
( 1 + W_p (\aeff_k)^{7-p} )^{1/2} }
\end{equation}
which depends on both  $l_i$, $i=1...n$, and the angles through
the function $W_p$ in \eqref{Np}.

If we consider the phase space parameterized by $N$, $\gym$, $r$,
$b$ and $l_i$, $i=1 \ldots n$ we can investigate, as done in
Section \ref{secphasestruc} for zero $l_i$, the phase structure
when varying $\zz = \geff^2$. From the metric \eqref{nhmet} and the
effective dilaton \eqref{effdilangmom} the deformations caused by
the presence of non-zero  angular momenta will depend on the
ratios \( l_i / r \), \( i=1 \ldots n\). If we  parametrize, along
with \eqref{phasepath},  the angular momenta as
\begin{equation}
\label{pathli} l_i \propto z^\beta \sp i=1 \ldots n
\end{equation}
then the deformations arising from the angular momenta do not
change with $\zz$.

The analysis of Section \ref{secphasestruc} used two special
calibration points, $\zz=1$ where the curvature of the geometry is
of order 1 and  $\zz=\xt$ where $\gseff \sim 1$. It might seem that
 we cannot define these points anymore, since for a specific
$\gym$, $b$ and $r$ we have a specific $\zz=\geff^2$, but both the
curvature of the geometry and the effective dilaton field are
clearly angular dependent. However,  considering for definiteness
the \( \xt \) point, we can define this instead by stating that
for $\zz \ll \xt$ we have $\gseff \ll 1$ and $\zz \gg \xt$ we have
$\gseff \gg 1$ (assuming that $\gseff$ is increasing near $\xt$).
In other words, we only have to define points on the phase
diagrams for $\zz$ up to a certain order, defined by the large
inequalities \( \gg \) and \( \ll \). Therefore, for a given set
of \( l_i / r \) we can choose the scale of the variables $N$,
$\gym$, $b$, $r$ and $l_i$, so that the phase transition points in
the phase diagram are defined with an accuracy good enough to have
regions with distinct phases. In this sense, the ratios \( l_i / r
\), \(i=1 \ldots n\) can be of any order, so long as we choose the
phase transition points to be of high enough order  for them to be
well defined. Hence, non-zero angular momenta do not induce any
modifications to the gauge coupling phase structure found in
Section \ref{secphasestruc}.

%------------------------------------
\section{Thermodynamics of NCSYM from supergravity \label{secnhtherm} }

\subsection{The thermodynamic quantities \label{secthermquan} }

In Section \ref{secdbrane} it was shown that the thermodynamics of
a spinning D-brane bound state is essentially  the same as that of
the spinning D$p$-brane, with $p$ being the spatial dimension of
the D-brane bound state. The only change is the appearance of
extra charges and  chemical potentials, but the temperature $T$,
entropy $S$, angular velocity $\Omega_i$ and angular momentum
$J_i$ were unchanged. We now consider the thermodynamics of the
near-horizon solution given in Section \ref{secnearhorsol}, which
can be obtained from \eqref{pbranethermo} using the rescalings
\eqref{rescal1}, \eqref{rescal2}. We also note that the energy is
computed from the energy above extremality $E=M- (\sum Q^2)^{1/2}$
in the near-horizon limit, where the sum is over all charges in
the bound state. Here, it is used that at extremality the bound
state is a 1/2-BPS state, and we recall that the expressions of
the charges and chemical potentials are given in Appendix
\ref{apprrpot}.

One then finds that  the charges as well as the chemical
potentials are constant in the near-horizon limit, and hence do
not appear in the thermodynamics. For the remaining thermodynamic
quantities we find in the near-horizon limit
\begin{subequations}
\label{nhthermdyn}
\begin{equation}
\label{nhtands}
T = \frac{w_p}{4\pi} \lambda^{-1/2} (7-p-2\kappa)
\frac{r_0^{\frac{7-p}{2}}}{r_H}
\sp
S = 4\pi \hat{w}_p V_p N^2 \lambda^{-3/2} r_0^{\frac{7-p}{2}} r_H
\end{equation}
\begin{equation}
\label{nhoandj}
\Omega_i = w_p \lambda^{-1/2} r_0^{\frac{7-p}{2}} \frac{l_i}{l_i^2+r_H^2}
\sp
J_i = 2 \hat{w}_p V_p N^2 \lambda^{-3/2} r_0^{\frac{7-p}{2}} l_i
\end{equation}
\begin{equation}
\label{nhenergy}
 E = \frac{9-p}{2} \tilde{w}_p V_p N^2 \lambda^{-2}
r_0^{7-p}
\end{equation}
\end{subequations}
where we have defined the NC 't Hooft coupling,
\begin{equation}
\label{nchooft}
 \lambda = \gym^2 N \prod_{k=1}^m b_k
\end{equation}
and where
\begin{subequations}
\begin{equation}
w_p = \sqrt{ \frac{(7-p)V(S^{8-p})}{(2\pi)^{9-2p}}}
\sp
\hat{w}_p = (2\pi)^{2p-11} \sqrt{\frac{ (2\pi)^{9-2p} V(S^{8-p})}{7-p} }
\end{equation}
\begin{equation}
\tilde{w}_p = w_p \hat{w}_p = (2\pi)^{2p-11} V(S^{8-p})
\end{equation}
\end{subequations}
It is not difficult to verify that the energy
satisfies the first law of thermodynamics
\begin{equation}
\label{nhfirstlaw}
 dE = T dS + \sum_{i=1}^n \Omega_i dJ_i \sp
E=E(S,\{ J_i \})
\end{equation}
and the integrated Smarr formula \cite{Harmark:1999xt}
\begin{equation}
(7-p) E = \frac{9-p}{2} TS + \frac{9-p}{2} \sum_{i=1}^n \Omega_i
J_i
\end{equation}

Comparing the thermodynamics \eqref{nhthermdyn} with the one obtained
for the near-horizon spinning D$p$-brane in Ref.
\cite{Harmark:1999xt} we observe that they are identical up to the
replacement
\begin{equation}
\gym^2 \rightarrow \gym^2 \prod_{k=1}^m b_k
\end{equation}
where $\gym$ on the left-hand side is the YM coupling constant of
the commutative theory and $\gym$ on the right-hand side of the
non-commutative theory. For the non-rotating case this was found
in
\cite{Maldacena:1999mh,Alishahiha:1999ci,Barbon:1999mx,Cai:1999aw}
and argued at weak coupling from the field theory point of view in
\cite{Bigatti:1999iz}.

For various applications it is also useful to compute the Gibbs
free energy $F$ from Eqs. \eqref{nhthermdyn}, with the result
\begin{equation}
\label{nhgibbs}
F = E-TS -\sum_{i=1}^n \Omega_i J_i = - \frac{5-p}{9-p} E
= - \frac{5-p}{2} \tilde{w}_p V_p N^2 \lambda^{-2} r_0^{7-p}
\end{equation}
satisfying
\begin{equation}
dF = - S dT - \sum_{i=1}^n J_i d\Omega_i
\sp
F = F(T,\{ \Omega_i \})
\end{equation}
For $p=2,3,4$ and one non-zero angular momentum the exact
form of the free energy in terms of the intensive thermodynamic
quantities is \cite{Harmark:1999xt}
\begin{subequations}\label{strongdpfe}
\begin{equation}
 F (T,\Omega) = - c_p V_p N^2
\lambda^{-\frac{p-3}{p-5}} T^{\frac{2(7-p)}{5-p}}
(1+x)^{2(6-p)/(5-p)} \left( 1 + \frac{x }{x_c}
\right)^{-2\frac{7-p}{5-p}}
\end{equation}
\begin{equation}
\label{xsol} x \equiv \frac{l^2}{r_H^2}
= 8 \left( \frac{7-p}{4 \pi} \right)^2
 \frac{\omega_c^4}{ \omega^2}
\left( 1 - \frac{1}{2} \left( \frac{ \omega}{
\omega_c} \right)^2 - \sqrt{ 1- \left( \frac{ \omega}{
\omega_c} \right)^2 } \right)
\sp \omega \equiv \frac{\Omega}{T}
\end{equation}
\end{subequations}
where the constants $c_p$ can be found in Ref. \cite{Harmark:1999xt}
and
\begin{equation}
\label{xc}
 x_c = \frac{7-p}{5-p} \sp \omega_c = \frac{2\pi}{\sqrt{(7-p)(5-p)}}
\end{equation}
The boundaries of stability and critical exponents for these cases
can also be found in Ref. \cite{Harmark:1999xt}, and we note in
particular the stability bound $2 \pi J/S \leq \sqrt{x_c}$, with
the boundary of   stability at $\omega = \omega_c$. For more than
one angular momentum the free energy has the same prefactor as in
\eqref{strongdpfe} and the expansion in powers of $\omega$
including the first two subleading terms can be found in
\cite{Harmark:1999xt}. For the case $p=5$ the free energy is zero
since $T$ and $\Omega_i$ are not independent variables anymore.
For the corresponding results in the D6-brane case we refer to
Section \ref{secd6}, which includes a more detailed account of the
thermodynamics as it shows some exotic behavior and will be
relevant for  the 7-dimensional non-commutative field theory that
is dual to this supergravity solution.

%----------------------------------
\subsection{Ranges of validity
\label{secrangeval} }

In this section we discuss the range of validity in which the
thermodynamics of the near-horizon  limit (see  Section
\ref{secthermquan}) describes the thermodynamics of the NCSYM 
on the world-volume of the D-brane.
In particular, we supplement the analysis of the gauge coupling
phase structure of Section \ref{secphasestruc} with further
details on the gauge coupling phase space relevant for the special
case in which the effective gauge coupling can vary independently
of the near-horizon intensive thermodynamic quantities (case d of Section
\ref{secphasestruc}).  Specifically, we explain the cases for
which i)  the coupling can go all the way to infinity, and ii)
finite $N$ is allowed, and still have a valid thermodynamic
description.

The criterion that we impose for invariant thermodynamics is that
intensive, scale-independent quantities like the temperature $T$
and the R-voltages $\Omega_i$ must remain fixed, while the
extensive quantities like the entropy $S$ and the R-charges $J_i$
are allowed  to change when all the extensive quantities change
uniformly. As discussed in Sections \ref{secphasestruc} and
\ref{secangmom}, for fixed $N$ the gauge theory phase space is
characterized by variables $\gym$, $r_0$, $l_i$, $b$, and the path
in the gauge theory phase space is parameterized by
\begin{equation}
\label{phsp}
 \gym^2 \propto z^\alpha \sp r_0 \propto z^\beta \sp
l_i \propto \zz^\beta \sp b \propto \zz^\gamma
\end{equation}
where \( \zz = \geff^2 = \gym^2 N b^m r_H^{p-3} \).
It is trivial to see that \( r_H \propto \zz^\beta \).
Note that we use
$r_H$ here rather than $r$ in $\geff$ since the validity of the D-brane
thermodynamics is considered. From the intensive quantities in
\eqref{nhthermdyn} it then follows that we need to keep
\( r_0^{5-p} \lambda^{-1} \) fixed, yielding the restriction
\begin{equation}
\label{restr}
\alpha + m\gamma = (5-p)\beta
\end{equation}
As shown in Section \ref{secphasestruc}, it follows from
\eqref{abc} and \eqref{restr}
that \( \beta = \frac{1}{2} \) and \( \eta = 1 + 2\gamma \). The
scaling of the extensive quantities is then proportional to \(
\zz^{\frac{p-3}{2}} \) under the variation of \( \zz  \).

We choose \( \gamma = 0 \) in the following, which seems most
natural considering that the position commutators for the
non-commutative coordinates are proportional to $b$
\cite{Maldacena:1999mh,Alishahiha:1999ci}. With this choice, \(
\eta = 1 \), so that the non-commutative sector is \( \zz \gg \xnc
\) while the commutative sector is \( \zz \ll \xnc \). In the
regime \( \xnc \ll 1 \), $N$ can be finite. As noted in Section
\ref{secphasestruc}, the condition \( \eta \geq \frac{7-p}{2m} \)
is fulfilled for $p=4$ with $m=2$ and $p=5,6$ with $m\geq 1$. In
these cases, if also \( \xnc \ll N^{\frac{4}{7-p}} \) then the
effective coupling \( \zz = \geff^2 \) can go all the way to
infinity with the D-brane thermodynamics being valid\footnote{For
the D3 and D5-brane this holds for all $m$, as explained in
Section \ref{secdualtheor}.} for the dual field theory, as
depicted in the phase diagram of Figure \ref{fignoxt}.

{} From the above considerations, we see that the most interesting
part of parameter space is the region where \( \xnc \ll 1 \),
since here both  $N$ can be finite, and in the indicated cases we
also have that the D-brane thermodynamics is valid for the dual
field theory for as large coupling as desired. It is interesting
to note that  the requirement \( \xnc \ll 1 \) can be seen as
demanding the non-commutativity effects to be as significant as
possible. We therefore find this region expressed in terms of the
intensive thermodynamic parameters $T$ and $\Omega_i$. Since
$\beta = \frac{1}{2}$ we have from \eqref{phsp},
\begin{equation}
\label{invquan}
r_0 = \hat{r}_0 \sqrt{\zz}
\sp
r_H = \hat{r}_H \sqrt{\zz}
\sp
l_i = \hat{l}_i \sqrt{\zz}
\end{equation}
with the parameters $\hat{r}_0$, $\hat{r}_H$ and $\hat{l}_i$
invariant under variation of $\zz$.
With this, the invariant part of $\lambda$
is $\hat{\lambda} = \hat{r}_H^{3-p}$.
Using the definition \eqref{aeffx} of
$\xnc$ it follows after some substitutions that,
\begin{equation}
\xnc \sim b^{-2} \hat{r}_H^{-4}
\end{equation}
so that
\begin{equation}
\xnc \ll 1 \ \Leftrightarrow
\ \hat{r}_H \gg b^{-1/2}
\end{equation}
This corresponds to the UV region, as can be seen by writing \( E
= \hat{E} \zz^{\frac{p-3}{2}} \), where the $\zz$-independent part
of the energy $\hat{E}$ is of the same order as $\hat{r}_0^{7-p}
\hat{r}_H^{2p-6}$ so that $\hat{E} \gg b^{-\frac{p+1}{2}}$ since
\( \hat{r}_0 \geq \hat{r}_H \). This is in agreement with the fact
that non-commutative effects are expected to be of significance in
the UV-region.

Substituting \eqref{invquan} in \eqref{nhthermdyn} we obtain
for the intensive thermodynamic quantities
\begin{equation}
\label{apprtando}
T \sim \hat{r}_H \left( \frac{\hat{r}_0}{\hat{r}_H} \right)^{\frac{7-p}{2}}
(7-p-2\kappa)
\sp
\Omega_i \sim \hat{r}_H^2
\left( \frac{\hat{r}_0}{\hat{r}_H} \right)^{\frac{7-p}{2}}
\frac{\hat{l}_i}{\hat{l}_i^2+\hat{r}_H^2}
\end{equation}
Provided we are away from the region in which
\( \kappa \) is near \( \frac{7-p}{2} \), we then have
\begin{equation}
\label{hightemp}
\xnc \ll 1 \ \Leftrightarrow
\ T \gg b^{-1/2} \left( \frac{\hat{r}_0}{\hat{r}_H} \right)^{\frac{7-p}{2}}
\end{equation}
showing that we are in the high-temperature region of the
\( (T,\{ \Omega_i \} ) \) phase space.
If we have that \( \hat{l}_i / \hat{r}_0 \) is of order one or less,
then \( \hat{r}_H \sim \hat{r}_0 \) and \eqref{hightemp} reduces to
\begin{equation}
\label{hightemp2}
\xnc \ll 1 \ \Leftrightarrow
\ T \gg b^{-1/2}
\end{equation}
For large angular momenta we have that
 \( \hat{r}_H \ll \hat{r}_0 \), in which case
even larger temperatures are necessary.

If $\kappa$ (see Eq. \eqref{horkappa}) is very near 
$\frac{7-p}{2}$ it follows from
\eqref{apprtando} that low temperatures are allowed as well. For
$\kappa = \frac{7-p}{2}$ we have $T=0$, but we can clearly still
have \( \hat{r}_H \gg b^{-1/2} \). It is not clear what the
significance of this special region of the $(T,\{\Omega_i\})$
phase space is, but we note that this region is also of interest
because the temperature can be zero with the other thermodynamic
quantities being non-zero.

%------------------------------------------------------------------------
\subsection{Thermodynamics of the D6-brane theory with $B$-field
\label{secd6}}

In this section we analyze the thermodynamics of the near-horizon
spinning D6-brane with a $B$-field, reviewing and adding to the
analysis of Ref. \cite{Harmark:1999xt}. This theory is of interest
since it has been shown \cite{Maldacena:1999mh,Alishahiha:1999ci}
that the D6-brane with non-zero $B$-field decouples from gravity
in the near-horizon limit.  
 This suggests the existence of a
consistent 7-dimensional NCSYM, dual to the near-horizon
D6-brane solution. We can thus obtain information about the
thermodynamics of this 7-dimensional NCSYM with R-voltage and
R-charge by considering the spinning near-horizon D6-brane. As we
shall see, the D6-brane theory has various exotic features
distinguishing it from the other D$p$-brane theories.

Substituting $p=6$ in the general formulae \eqref{nhthermdyn} and
\eqref{nhgibbs} one obtains the thermodynamic quantities
\begin{subequations}
\label{d6quan}
\begin{equation}
\label{d6tands}
T = \sqrt{2} \pi \lambda^{-1/2} \frac{1-l^2/r_H^2}{\sqrt{r_0}}
\sp
S = 4 \sqrt{2} \pi V_6 N^2 \lambda^{-3/2} \sqrt{r_0} r_H
\end{equation}
\begin{equation}
\label{d6oandj}
\Omega = 4 \sqrt{2} \pi^2 \lambda^{-1/2} \frac{l}{r_H} \frac{1}{\sqrt{r_0}}
\sp
J = 2\sqrt{2} V_6 N^2 \lambda^{-3/2} \sqrt{r_0} l
\end{equation}
\begin{equation}
\label{d6eandf}
E = 12 \pi^2 V_6 N^2 \lambda^{-2} r_0
\sp
F = 4\pi^2 V_ 6 N^2 \lambda^{-2} r_0
\end{equation}
\end{subequations}
where the horizon radius is determined by \eqref{horkappa} as
\begin{equation}
r_H = \frac{r_0}{2}
\left( 1 + \sqrt{1- \frac{4l^2}{r_0^2} } \right)
\end{equation}
This means that the angular momentum and horizon radius are
restricted to the ranges \( 0 \leq l \leq r_0 /2 \) and \( r_0 /2
\leq r_H \leq r_0 \) respectively. From this we see that \( l \leq
r_H \) so that it follows from \eqref{d6tands} that \( T \geq 0
\) and that \( T = 0\) for \( l = r_0 / 2 = r_H \).

We now address the question of thermodynamic stability of the
D6-brane theory. We begin by considering the grand canonical
ensemble, in which the system is in contact with a heat reservoir
of temperature $T$ and R-voltage $\Omega$. Defining  \( \tilde{S}
= \sqrt{r_0} r_H \), \( \tilde{J} = \sqrt{r_0} l \) and \(
\tilde{E} = r_0 \), which are rescaled versions of $S$, $J$ and
$E$ in \eqref{d6quan}, we find the relation
\begin{equation}
\label{tildeS}
\tilde{S} = \frac{1}{2} \left( \tilde{E}^{3/2}
+ \sqrt{ \tilde{E}^3 - 4 \tilde{J}^2} \right)
\end{equation}
From \eqref{tildeS} one can check that the Hessian of the entropy
always has one negative and one positive eigenvalue, and the
theory is therefore thermodynamically unstable in the entire
$(E,J)$ phase space.

Another way to arrive at this result is by considering the Gibbs
free energy
\begin{equation}
F(T,\Omega) = 8 \pi^4 V_6 N^2 \lambda^{-3} T^{-2}
\xi \Big( (2\pi)^{-2} \omega^2 \Big)
\end{equation}
where \( \omega = \Omega / T \) and \( \xi \) is the function defined by
\begin{equation}
\xi (x) = -\frac{8}{x^2} \Big( \sqrt{1+x}-1-\frac{1}{2}x \Big)
= 1 - \frac{1}{2}x + \frac{5}{16} x^2 - \frac{7}{32} x^3 + \Ord (x^4 )
\end{equation}
for \( x \geq -1 \). Note that the Gibbs free energy is positive,
contrary to the Gibbs free energies in \eqref{strongdpfe} for the
other near-horizon D$p$-brane theories. The Gibbs free energy is a
function on the $(T,\Omega)$ phase space, the properties of which
depend on the map between the supergravity variables $(r_H,l)$ and
$(T,\Omega)$. For the D6-brane this map is one-to-one, which is
 a consequence of the fact that the determinant of the
Hessian of the Gibbs free energy \cite{Harmark:1999xt}
\begin{equation}
\det \hes (F) = - 6 \pi^{-2} V_6^2 N^4 \lambda^{-2} r_0^4
\left( 1 + \frac{l^2}{r_H^2} \right)^{-3}
\end{equation}
is neither zero nor singular. Since this determinant is always
negative, the theory is clearly unstable for all points in the
$(T,\Omega)$ phase space. We remark that because the map between
$(r_H,l)$ (or equivalently $(S,J)$) and $(T,\Omega)$ is
one-to-one, the phase mixing mechanism argued by Cvetic and Gubser
in \cite{Cvetic:1999rb} for the D3-brane, is impossible for the
 D6-brane theory.

Turning instead to the canonical ensemble, in which case the
system is in contact with a heat bath of temperature $T$ at a
fixed R-charge $J$, the heat capacity takes the form
\cite{Harmark:1999xt}
\begin{equation}
\label{CJ}
C_J = 12 \sqrt{2} \pi V_6 N^2 \lambda^{-3/2} \sqrt{r_0} r_H
\frac{(r_H^2-l^2) (r_H^2+l^2) }{5 l^4 + 8 l^2 r_H^2 - r_H^4}
\end{equation}
From this expression it follows that $C_J$ is negative for $0 \leq
l^2/r_H^2
< \frac{\sqrt{21}-4}{5} $, zero at $l^2/r_H^2 =
\frac{\sqrt{21}-4}{5} $, positive for $\frac{\sqrt{21}-4}{5} <
l^2/r_H^2 < 1 $ and zero again at $l^2/r_H^2 = 1$. The D6-brane
theory thus has the remarkable property that it is
thermodynamically stable in the canonical ensemble, in the range
$\frac{\sqrt{21}-4}{5} \leq l^2/r_H^2 \leq 1 $
\cite{Harmark:1999xt}. This can also be written as \(
\omega = \Omega/T \geq \omega_c \) with \( \omega_c \simeq 4.8551
\). In \cite{Harmark:1999xt} the weakly-coupled D6-brane theory
was studied in the ideal gas approximation where it was found to
exhibit  a qualitatively different behaviour, with the brane being
stable when \( \omega \leq \omega_c' \simeq 4.9948 \). The weak
and strong coupling limits of this theory should therefore be
connected via a rather non-trivial phase transition.

If we consider the Euclidean version of the D6-brane theory, we
have to perform the Wick rotation \( \tau = it \) and \( \tilde{l}
= -il \). This yields the restriction \( 0 \leq \tilde{l}^2/r_H^2
< 1 \) for the Euclidean theory, since \( r_0 =
(1-\tilde{l}^2/r_H^2) r_H \). When \( \tilde{l}^2/r_H^2
\rightarrow 1 \)  the supersymmetry of the theory is restored,
since the supersymmetric boundary conditions are recovered from
the R-symmetry group element in the partition function
\cite{Cvetic:1999rb}.

The corresponding expression for the entropy is obtained with the
substitution \( \tilde{J}^2 \rightarrow -\tilde{J}^2 \) in
\eqref{tildeS}, and it is easy to check that the Hessian of the
entropy always has two negative eigenvalues. Thus, the Euclidean
D6-brane theory is also thermodynamically unstable in the $(E,J)$
phase space, so there are no stable regions in the grand canonical
ensemble. If we consider the canonical ensemble, it is easily seen
that the heat capacity \eqref{CJ} is never positive, and hence
there are no stable regions in this ensemble either. The limit  \(
\tilde{l}^2/r_H^2 \rightarrow 1 \) appears  to be singular as, for
example, the temperature diverges in this limit. The issue of
stability is not clear, since the second derivatives of the
entropy or the free energies go to zero or infinity. It is thus
unclear, and highly questionable, whether a sensible theory can be
recovered in this limit.

%----------------------------------------
\section{Spinning M-brane bound states and non-commutative $(2,0)$
theory \label{secm5} }

\subsection{Spinning M5-M2 brane bound state}

 The spinning M5-M2 brane bound state can be obtained from the spinning
 D4-D2 brane bound state ($p=4, m=1$) of Section \ref{secdbrane} by
lifting to M-theory, which gives the metric
\begin{eqnarray}
\label{solmetm5} ds^2 &=& (H D)^{-1/3} \left[  - f_5 dt^2
+(dy^1)^2 + (dy^2)^2 + D\Big( (dy^3)^2 + (dy^4)^2 + (dy^5)^2
\Big)\right. \nn \\ & & + H
 \Big( \bar{f}_5^{-1} K_5 dr^2 + \Lambda_{\alpha
\beta} d\eta^\alpha d\eta^\beta \Big) \nn \\ && \left. +
\frac{1}{K_5 L_5} \frac{r_0^{3}}{r^{3}} \Big( \sum_{i,j=1}^2 l_i l_j
\mu_i^2 \mu_j^2 d\phi_i d\phi_j - 2 \cosh \alpha \sum_{i=1}^2 l_i
\mu_i^2 dt d\phi_i \Big) \right]
\end{eqnarray}
and gauge potentials
\begin{subequations}
\label{solpotm5}
\begin{equation} {\cal{C}}_3
= -\sin \theta (H^{-1}-1) I \wedge dy^1 \wedge dy^2 + \tan \theta
D H^{-1} dy^3 \wedge dy^4 \wedge dy^5
\end{equation}
\begin{equation}
{\cal{E}}_6 = \cos \theta D (H^{-1} -1) I \wedge dy^1 \wedge
\cdots \wedge dy^5
\end{equation}
\end{subequations}
Here, the functions $f_5$, $\bar f_5$, $K_5$ and $L_5$ are
as in \eqref{defs} with $p=4$, $D$ is defined in
\eqref{ddef} and the one-form $I$ is defined in
\eqref{Idef}. We also need the relations
\begin{equation}
\label{M5rel}
 16 \pi G = (2 \pi)^8 l_p^9 \sp h^3 =
\frac{\hat{h}^3}{\cos \theta} \sp  \hat{h}^3 = \pi N l_p^3
\end{equation}
where $l_p$ is the 11-dimensional Planck length, $h$ is defined in
\eqref{hdef} (with $p=4$) and the second relation follows from
charge quantization of the M5-brane. For zero angular momentum the
solution reduces to the one given in Ref. \cite{Russo:1997if}.

Aside from the presence of charges and chemical potentials for both
the M5 and M2-brane,
the thermodynamic quantities of this background are not altered due
to the presence of $D$, and coincide with those of the spinning M5-brane.
For example, it is not
difficult to see that the temperature is not modified by the presence of
$D$ and also cancels out in the horizon area (and hence the entropy); the
 same conclusion holds for the mass, angular momentum and velocity.
The complete set of thermodynamic quantities is then given by
\begin{subequations}
\label{thermm5}
\begin{equation}
M = \frac{V_5 V(S^{4})}{16 \pi G} r_0^{3}
\Big[ 4 + 3 \sinh^2 \alpha \Big]
\end{equation}
\begin{equation}
\label{tandsm5} T = \frac{3-2\kappa }{4 \pi r_H \cosh \alpha } \sp S
= \frac{V_5 V(S^{4}) }{4G} r_0^{3} r_H \cosh \alpha 
\end{equation}
\begin{equation}
\label{angmomm5} \Omega_i = \frac{l_i}{(l_i^2 + r_H^2)\cosh \alpha }
\sp J_i = \frac{V_5 V(S^{4})}{8\pi G} r_0^{3} l_i \cosh \alpha 
\end{equation}
\begin{equation}
\label{mum5} \mu_5 = \cos \theta \mu \sp \mu_2 = -\sin \theta \mu
\sp  Q_5 = \cos \theta Q \sp
 Q_2 = -\sin \theta Q
\end{equation}
\begin{equation}
\mu =  \tanh \alpha \sp   Q = \frac{V_5 V(S^{4})}{16\pi G} 3
r_0^{3}  \sinh \alpha  \cosh \alpha 
\end{equation}
\end{subequations}
satisfying the integrated Smarr formula
\begin{equation}
3 M = 4 TS + 3 (\mu_5 Q_5 + \mu_2 Q_2) + 4 \sum_{i=1}^2 \Omega_i
J_i
\end{equation}
which is a consequence of the first law of thermodynamics
\eqref{firstlaw}.

\subsection{The near-horizon limit}
Next we turn to the near-horizon limit of the spinning M5-M2 brane
bound state, which corresponds
\cite{Maldacena:1999mh,Seiberg:1999vs,Alishahiha:1999ci} to the six-dimensional
non-commutative (2,0) theory \cite{Seiberg:1997zk}.  
The near-horizon limit is
defined by letting the Planck length $l_p \rightarrow 0 $
accompanied by the rescalings\footnote{Note that $r$ corresponds
to energy squared in our conventions.}
\begin{subequations}\label{rescal}
\begin{eqnarray}
 && r  = \frac{\old{r}}{l_p^3}   \sp r_0  =
\frac{\oldp{r_0}}{l_p^3}  \sp l_i = \frac{\oldp{l_i}}{l_p^3} \\ &
&  \hat h^{3} = \frac{\old{\hat h}^{3}}{l_p^3} \sp c  = l_p^6 \tan
\theta \\ &&  y_i = \frac{l_p^3}{c^{1/2}} \oldp{y_i}  \;\; ,
i=0,1,2 \sp y_i =  \frac{c^{1/2}}{l_p^3} \oldp{y_i} \;\; ,
i=3,4,5
  \\ \label{solrescal}
 && ds^2 = \frac{\old{ds^2}}{l_p^2}
 \sp \C_3 = \frac{\oldp{\C_3}}{l_p^3} \sp
  \E_6 = \frac{\oldp{\E_6}}{l_p^6} \sp
   G  = \frac{\old{G}}{l_p^9}
\end{eqnarray}
\end{subequations} 
 where  the quantities on the left hand side are kept fixed
 and the quantities on the right-hand side (labelled with subscript
  ``old'' except for $\theta$) are those that enter the
  asymptotically-flat solution \eqref{solmetm5}, \eqref{solpotm5}.
  Note that  the rescaling in
\eqref{solrescal} leaves the 11-dimensional low-energy
supergravity action  invariant. It also follows that $e^{2 \alpha }
\rightarrow 4 c(\hat h/r_0)^3 l_p^{-12}$, which is used below.

The metric of the resulting near-horizon background of the
spinning M5-brane with non-zero $\C_3$-field is then given by
\begin{eqnarray}
\label{solm5nh}
 ds^2 &=& (H D)^{-1/3} \left[  - f_5 dt^2
+(dy^1)^2 + (dy^2)^2 + D[ (dy^3)^2 + (dy^4)^2 + (dy^5)^2
]\right. \nn
\\ & & + H
 \Big( \bar{f}_5^{-1} K_5 dr^2 + \Lambda_{\alpha
\beta} d\eta^\alpha d\eta^\beta \Big) \left. - 2 \frac{1}{K_5 L_5}
\frac{(r_0 \hat h)^{3/2}}{r^{3}} \sum_{i=1}^2 l_i \mu_i^2 dt
d\phi_i  \right]
\end{eqnarray}
where
\begin{equation}
H = \frac{\hat h^3}{K_5 L_5 r^3} \sp D = [ 1 + a^3 K_5 L_5 r^3]^{-1} \sp
a^3 = \frac{c}{\hat h^3}
\end{equation}
and $\hat h^3 = \pi N$ after the rescaling.
 The expressions
for the rescaled 3-form potential is likewise given by
\begin{equation}
\C_3 =  -c \left( H^{-1}  d t + \left( \frac{r_0}{\hat h}
\right)^{3/2} \sum_{i=1}^2 l_i \mu_i^2 d \phi_i \right)
 \wedge dy^1 \wedge dy^2 + \frac{1}{c^2} H^{-1}  D
  dy^3 \wedge dy^4 \wedge dy^5
\end{equation}
where we have ignored a constant term that does not contribute to
the field strength. The expression for the potential $\E^{\rm
tot}_6=\E_6- \frac{1}{2}\C_3^2$ entering the 7-form field strength
can be obtained in the same way, but is a bit more involved.

To compute the thermodynamic quantities in the near-horizon limit,
one may use the quantities  \eqref{thermm5} of the
asymptotically-flat solution along with the rescaling
\eqref{rescal}, and employ the formula $E= M- (\sum Q^2)^{1/2}$
for the internal energy. However, since the time coordinate is
also rescaled,  we also have that
\begin{equation}
E = \old{E} \frac{c^{1/2}}{l_p^3} \sp T = \old{T}
\frac{c^{1/2}}{l_p^3} \sp \Omega_i = \oldp{\Omega_i}
\frac{c^{1/2}}{l_p^3}
\end{equation}
so that the following thermodynamics is obtained in the
near-horizon limit
\begin{subequations}
\label{thermnhm5}
\begin{equation}
E = \frac{5}{3} \frac{V_5}{(2 \pi)^6} r_0^{3}
\end{equation}
\begin{equation}
\label{tandsnh5} T = \frac{3-2\kappa }{4 \pi r_H }
\frac{r_0^{3/2}}{(\pi N)^{1/2}}
 \sp S = \frac{4}{3}  \frac{V_5(\pi N)^{1/2}}{(2 \pi)^5} r_0^{3/2} r_H
\end{equation}
\begin{equation}
\label{angmommnh5} \Omega_i = \frac{l_i}{(l_i^2 + r_H^2) }
\frac{r_0^{3/2}}{ (\pi N)^{1/2}}
 \sp J_i = \frac{4}{3}
\frac{V_5 (\pi N)^{1/2}}{(2 \pi)^6 } r_0^{3/2} l_i
\end{equation}
\end{subequations}
satisfying the near-horizon Smarr formula \cite{Harmark:1999xt}
\begin{equation}
\label{smarrnh} 3 E = \frac{5}{2} TS + \frac{5}{2} \sum_{i=1}^2
\Omega_i J_i
\end{equation}
which is a consequence of the first law of thermodynamics
\eqref{nhfirstlaw}.

Eq. \eqref{thermnhm5} describes the thermodynamics of the
non-commutative (2,0) theory, and is manifestly  independent of
$c$, i.e. coincides with that of commutative (2,0) theory. Just as
for the D$p$-brane case, the only difference of the presence of
the non-zero $\C_3$ potential is that the range of validity can be
different now. In fact, focusing on zero angular momentum first,
the curvature is \cite{Alishahiha:1999ci}
\begin{equation}
R \sim \left( N^{2/3} (1 + a^3 r^3) \right)^{-1}
\end{equation}
which needs to be small in order to trust the supergravity
description. This can be achieved either in the limit $N \gg 1$ as
in the commutative case, but now there is the additional
possibility of keeping $N$ finite and requiring $r \gg 1/a$. This
shows that the larger the non-commutative parameter $a$, the
larger the energy range in which the near-horizon limit is a valid
description of the non-commutative (2,0) theory. Following a
similar reasoning as in Section \ref{secangmom} this conclusion
holds also when the angular momenta are non-zero.

Finally, the general formulae of Ref. \cite{Harmark:1999xt} may be
used to compute e.g. the internal energy and Gibbs free energy in
terms of the extensive or intensive thermodynamic quantities
 respectively. For example, for one non-zero angular
momentum the free energy takes the form
\begin{equation}
F (T,\Omega)  = -\frac{2^6 \pi^3}{3^7} N^3  V_5 T^6 (1+x)^{4}
\left( 1 + \frac{x }{x_c} \right)^{-6}
\end{equation}
where $x$, $x_c$ are given by the $p=4$ expressions in
\eqref{xsol}, \eqref{xc}.
 For the resulting  boundaries of stability and critical exponents
of the spinning M5-brane see
Refs. \cite{Cai:1998ji,Cvetic:1999rb,Harmark:1999xt}.

\section{Conclusions and discussion \label{seccon}}

In this paper we have constructed general spinning brane bound
state solutions of string and M-theory, and discussed their
thermodynamics, extending our work \cite{Harmark:1999xt}. Except
for additional charges and chemical potentials of the lower branes
in the bound state, the thermodynamics is equivalent to that of
the highest brane in the bound state. We have computed the
near-horizon limits of these supergravity solutions, which are
dual to NCSYM theories for string theory and the non-commutative (2,0)
theory for M-theory. Using this correspondence, we have
presented a general analysis of the gauge coupling phase structure
of these NCSYM theories by considering a general path in their
phase space, both for zero as well as non-zero angular momenta.
This analysis, which includes the one in Ref. \cite{Alishahiha:1999ci} as a
special case, exhibits various interesting features, including
regions in which the supergravity description is valid for finite
$N$ and/or  the effective coupling ranging from the transition point
$\geff \sim 1$
all the way to infinity. More generally, four  types of phase
diagrams are found, and we have established that for each spatial
worldvolume dimension of the brane and each non-zero rank of
$B$-field, a path and region of phase space can be chosen such
that the phase structure of any of the four phase diagrams can be
realized.

The thermodynamics of the near-horizon solutions is not altered by
the presence of the $B$-field, in parallel with results for the
non-rotating case
\cite{Maldacena:1999mh,Alishahiha:1999ci,Barbon:1999mx,Cai:1999aw}
showing that, to leading order, the thermodynamics of SYM is
equivalent to that of NCSYM. This was argued at weak coupling from
the field theory point of view \cite{Bigatti:1999iz} by showing
that the planar limit of SYM and NCSYM coincide. As an application
of the general phase structure analysis, the validity of the
thermodynamics for the NCSYM has been examined by requiring that
the intensive thermodynamic parameters are invariant for the path
in phase space. We have determined the region of phase space in
which $N$ can be finite and at the same time the coupling can be
taken all the way to infinity. The resulting condition is that
\begin{equation}
\label{hightemp3} \ T \gg b^{-1/2} \left(
\frac{\hat{r}_0}{\hat{r}_H} \right)^{\frac{7-p}{2}}
\end{equation}
showing that that at fixed non-extremality parameter $\hat r_0$
and horizon radius $\hat r_H$, the larger the non-commutativity
parameter $b$, the larger the temperature region in which these
properties are satisfied. Interestingly, for $\kappa$ near
$\frac{7-p}{2}$ low temperatures are allowed as well. 

Having non-zero angular momenta does not qualitatively change the
gauge coupling phase structure, but may well provide further
insights into  NCSYM in the presence of voltages for the
R-charges. Moreover, as another application we have discussed the
D6-brane theory in further detail, adding to the results of
\cite{Harmark:1999xt}. This theory is of interest in view of the
recent discovery that the D6-brane theory with $B$-field decouples
from gravity \cite{Maldacena:1999mh,Alishahiha:1999ci}. Moreover,
while the non-rotating case
 is thermodynamically unstable, for the spinning D6-brane stability
is found in the canonical ensemble for sufficiently high angular
momentum density\footnote{We note that the D6-brane theory is also 
related to M(atrix) theory  on $T^6$ 
 \cite{Losev:1997hx}.}.  

Unlike in the commutative case, the non-commutative setup allows
for situations in which the supergravity description is valid for
finite $N$. 
While in the usual large $N$ limit the $1/N$
corrections\footnote{See Ref. \cite{Barbon:1999mx} for a
discussion of $1/N$ corrections to the entropy of NCSYM.} around
the planar limit are generated by the string loop expansion, for
finite $N$ this cannot be true anymore. Indeed, it is not
difficult to see that in this case the effective string coupling
\eqref{gseff} is small not because $N$ is large, but rather since
the effective NC parameter $ \aeff $ is large, so that the string
loop expansion becomes an expansion in $1/ \aeff$. This indicates
that, in some sense, the framework enables one to interchange the
large $N$ expansion with a large $\aeff$ expansion. We also note
that the $\alpha'$ expansion of higher derivative terms in the
effective action, will generate $1/\lambda$
corrections\footnote{For the D3-brane case, tree-level $R^4$
corrections have been recently addressed in \cite{Cai:1999aw}.},
with $\lambda$ the NC 't Hooft coupling \eqref{nchooft}.

The argument above provides insight into the issue raised in 
section \ref{secphasestruc}: While the supergravity
description for SYM in the strong coupling and large $N$ limit 
gives extensive thermodynamic quantities proportional
to $N^2$, at weak coupling and finite $N$ 
these scale with $N^2-1$, coming from the $SU(N)$ group. In the
commutative AdS/CFT correspondence for $N$ D3-branes it is
believed \cite{Green:1999qt} that string loop corrections scaling as $1/N^2$,
will generate the correct $N^2-1$ factor, in accord with
strong/weak coupling duality of N=4 SYM. 
On the other hand, for strongly coupled NCSYM at finite $N$ (and thus
large $\aeff$) the extensive thermodynamic quantities are proportional 
to $N^2$. 
However, as argued above, in this case the
$1/\aeff$ corrections should generate the desired correction to
$N^2$ to obtain $N^2-1$, thereby connecting NCSYM to SYM. 
We finally note that the $N^2$ dependence  is in agreement with 
the observation that the non-commutative field theory at weak coupling 
 has gauge group  $U(N)$ \cite{Seiberg:1999vs} in order for
the group generators to form a closed algebra under matrix multiplication.

\section*{Note added}  
After completion of this work the interchange of large $N$ with large $\aeff$
expansion was also found at weak coupling by showing that for large $\aeff$
only the planar diagrams survive \cite{Minwalla:1999px}.

\section*{Acknowledgments}
We thank
J. Correia, P. Di Vecchia, R. Marotta, J.L. Petersen, K. Savvidis
and R. Szabo for useful discussions.
This work is supported in part by TMR network ERBFMRXCT96-0045.

%----------------------------------------

\begin{appendix}

\section{T-duality as a solution generator \label{apptdual}}

The spinning bound state solution \eqref{solmet}-\eqref{solb} (and
RR potentials given in Appendix \ref{apprrpot}) can be obtained
from the general spinning D$p$-brane solution of
\cite{Harmark:1999xt} by repeated use of the following sequence of
T-dualities and coordinate transformations
\cite{Russo:1997if,Breckenridge:1997tt}. For each disjoint pair of
spatial dimensions in the world-volume of the D$p$-brane the
relevant part of the background is
\begin{subequations}
\begin{equation}
ds^2 = \sum_{i,j=1}^2 g_{ij} dy^i dy^j \sp g_{11} = g_{22} =
H^{-1/2} \sp g_{12} = 0
\end{equation}
\begin{equation}
B_{12} = 0 \sp e^{2\phi} = H^{(3-p)/2}
\end{equation}
\end{subequations}
Performing a T-duality transformation in the \( y^2 \)-direction
we obtain
\begin{subequations}
\begin{equation}
ds^2 = \sum_{i,j=1}^2 {g'}_{ij} dy^i dy^j
\sp
{g'}_{11} = H^{-1/2}
\sp
{g'}_{22} = H^{1/2}
\end{equation}
\begin{equation}
{B'}_{12} = 0
\sp
e^{2{\phi'}} = e^{2\phi} H^{1/2}
\end{equation}
\end{subequations}
Next, one rotates the coordinates
\begin{equation}
\label{rotation}
 \vecto{\tilde{y}^1}{\tilde{y}^2} = \matrto{\cos
\theta}{\sin \theta}{-\sin \theta}{\cos \theta} \vecto{y^1}{y^2}
\end{equation}
yielding the background
\begin{subequations}
\begin{equation} ds^2 = \sum_{i,j=1}^2 \tilde{g}_{ij}
d\tilde{y}^i d\tilde{y}^j
\end{equation}
\begin{eqnarray}
\tilde{g}_{11} &=& H^{-1/2} \cos^2 \theta + H^{1/2} \sin^2 \theta
\\
\label{g22} \tilde{g}_{22} &=& H^{-1/2} \sin^2 \theta + H^{1/2}
\cos^2 \theta = H^{1/2} D^{-1}
\\
\label{g12}
\tilde{g}_{12} &=& \sin \theta \cos \theta \Big(
H^{1/2} - H^{-1/2} \Big)
\end{eqnarray}
\begin{equation}
\tilde{B}_{12} = 0
\sp
e^{2\tilde{\phi}} = e^{2\phi} H^{1/2}
\end{equation}
\end{subequations}
where
\begin{equation}
\label{didef}
 D^{-1} = \cos^2 \theta + \sin^2 \theta H^{-1}
\end{equation}
Finally, we make a T-duality transformation in \( \tilde{y}^2 \)
and we obtain
\begin{subequations}
\begin{equation}
\hat{g}_{11} = \frac{1}{\tilde{g}_{22}} \Big( \tilde{g}_{11} \tilde{g}_{22}
- \tilde{g}_{12}^2 \Big) - \frac{\tilde{B}_{12}^2}{\tilde{g}_{22}}
= \frac{1}{\tilde{g}_{22}} = H^{-1/2} D
\end{equation}
\begin{equation}
\hat{g}_{22} = \frac{1}{\tilde{g}_{22}} = H^{-1/2} D
\sp
\hat{g}_{12} = - \frac{\tilde{B}_{12}}{\tilde{g}_{22}} = 0
\end{equation}
\begin{equation}
\hat{B}_{12} = - \frac{\tilde{g}_{12}}{\tilde{g}_{22}}
= \sin \theta \cos \theta \Big( H^{-1} - 1 \Big) D
= \tan \theta \Big( H^{-1} D - 1 \Big)
\end{equation}
\begin{equation}
e^{2\hat{\phi}} = e^{2\tilde{\phi}} \frac{1}{\tilde{g}_{22}}
= e^{2\phi} D
\end{equation}
\end{subequations}

For the RR fields the T-duality transformations are in the most
general case (with repeated application of the above prescription
for disjoint pairs of spatial world-volume coordinates) a bit more
involved \cite{Bergshoeff:1995as}, but as an illustration we work
out the first application on coordinates $y^1$ and $y^2$. We start
with the RR-field \( A_{t12\cdots p} \) and T-duality in \( y^2 \)
gives
\begin{equation}
A'_{t13\cdots p} = - A_{t13\cdots p2}
\end{equation}
Applying the  rotation \eqref{rotation} one obtains
\begin{subequations}
\begin{equation} \tilde{A}_{t13\cdots p} = \cos \theta
A'_{t13\cdots p} = -\cos \theta A_{t13\cdots p2}
\end{equation}
\begin{equation}
\tilde{A}_{t23\cdots p} = - \sin \theta A'_{t13\cdots p}
= \sin \theta A_{t13\cdots p2}
\end{equation}
\end{subequations}
so that the final T-duality in \( \tilde{y}^2 \) gives
\begin{subequations}
\begin{equation}
\hat{A}_{t3\cdots p} = - \tilde{A}_{t3\cdots p2}
= -(-1)^p \tilde{A}_{t2\cdots p}
= -(-1)^p \sin \theta A_{t13\cdots p2}
= -\sin \theta A_{t12\cdots p}
\end{equation}
\begin{eqnarray}
&& \hat{A}_{t12\cdots p} = (-1)^{p} \hat{A}_{t13\cdots p2} =
(-1)^{p} \left[ -\tilde{A}_{t13\cdots p} + \tilde{A}_{t23\cdots p}
\frac{\tilde{g}_{12}}{\tilde{g}_{22}} \right] \nn \\ && =\Big(
\cos \theta + \sin \theta \frac{\tilde{g}_{12}}{\tilde{g}_{22}}
\Big) A_{t12\cdots p} = \cos \theta D A_{t12\cdots p}
\end{eqnarray}
\end{subequations}
where we used \eqref{g12},\eqref{g22} and the definition
\eqref{didef}.

\section{RR gauge potentials \label{apprrpot}}

In this appendix we give the RR potentials of the
asymptotically-flat spinning bound state solution
\eqref{solmet}-\eqref{solb}, which can be obtained from the
procedure outlined in Appendix \ref{apptdual} using in particular
the T-duality transformations of the RR fields given in Ref.
\cite{Bergshoeff:1995as}. To write these expressions we define the
following one-form, relevant for spinning brane solutions
%\cite{Harmark:1999xt}
%
\begin{equation}
\label{Idef}
 I = \frac{1}{\sinh \alpha } \Big( \cosh \alpha dt -
\sum_{i=1}^n l_i \mu_i^2 d\phi_i \Big)
\end{equation}
and we consider  maximal rank $m=[p/2]$, since lower rank can be
obtained by setting the appropriate $\theta_k$ to zero.

 The RR fields for the cases $p=2 \ldots 6$ are then given
by the following expressions: For the $p=2$ (D2-D0) case we have
\begin{equation} A_1 = - \sin \theta_1 \Big( H^{-1} -1 \Big)
I \sp A_3 = \cos \theta_1 D_1 \Big( H^{-1} -1 \Big) I \wedge dy^1
\wedge dy^2
\end{equation}
and for $p=3$ (D3-D1) analogously
\begin{equation}
A_2 = \sin \theta_1 \Big( H^{-1} -1 \Big) I \wedge dy^3
\sp
A_4 = - \cos \theta_1 D_1 \Big( H^{-1} -1 \Big) I \wedge dy^1 \wedge dy^2
\wedge dy^3
\end{equation}
The next case is $p=4$ (D4-D2-D0) with RR-fields
\begin{subequations}
\begin{equation}
A_1 = \sin \theta_1 \sin \theta_2 \Big( H^{-1} -1 \Big) I
\end{equation}
\begin{equation}
A_3 = - \Big( H^{-1}-1 \Big) I \wedge
\Big[ \cos \theta_1 D_1 \sin \theta_2 dy^1 \wedge dy^2
+ \sin \theta_1 \cos \theta_2 D_2 dy^3 \wedge dy^4 \Big]
\end{equation}
\begin{equation}
A_5 = \cos \theta_1 D_1 \cos \theta_2 D_2 \Big( H^{-1} -1 \Big)
I \wedge dy^1 \wedge \cdots \wedge dy^4
\end{equation}
\end{subequations}
 and analogously $p=5$ (D5-D3-D1)
\begin{subequations}
\begin{equation}
A_2 = -\sin \theta_1 \sin \theta_2 \Big( H^{-1} -1 \Big) I \wedge dy^5
\end{equation}
\begin{equation}
A_4 = \Big( H^{-1}-1 \Big) I \wedge
\Big[ \cos \theta_1 D_1 \sin \theta_2 dy^1 \wedge dy^2
+ \sin \theta_1 \cos \theta_2 D_2 dy^3 \wedge dy^4 \Big] \wedge dy^5
\end{equation}
\begin{equation}
A_6 = -\cos \theta_1 D_1 \cos \theta_2 D_2 \Big( H^{-1} -1 \Big)
I \wedge dy^1 \wedge \cdots \wedge dy^5
\end{equation}
\end{subequations}
Finally, for $p=6$ (D6-D4-D2-D0) the RR fields are
\begin{subequations}
\begin{equation} A_1 = - \sin \theta_1 \sin \theta_2 \sin
\theta_3 \Big( H^{-1}-1 \Big) I
\end{equation}
\begin{eqnarray}
A_3 &=& \Big( H^{-1}-1 \Big) I \wedge \Big[
\cos \theta_1 D_1 \sin \theta_2 \sin \theta_3 dy^1 \wedge dy^2
\nn \\ &&
+ \sin \theta_1 \cos \theta_2 D_2 \sin \theta_3 dy^3 \wedge dy^4
+ \sin \theta_1 \sin \theta_2 \cos \theta_3 D_3 dy^5 \wedge dy^6 \Big]
\end{eqnarray}
\begin{eqnarray}
A_5 &=& - \Big( H^{-1}-1 \Big) I \wedge \Big[
\sin \theta_1 \cos \theta_2 D_2 \cos \theta_3 D_3 dy^3 \wedge \cdots \wedge
dy^6
\nn \\ &&
+\cos \theta_1 D_1 \sin \theta_2 \cos \theta_3 D_3 dy^1 \wedge dy^2
\wedge dy^5 \wedge dy^6
\nn \\ &&
+\cos \theta_1 D_1 \cos \theta_2 D_2 \sin \theta_3 dy^1 \wedge \cdots \wedge
dy^4 \Big]
\end{eqnarray}
\begin{equation}
A_7 = \cos \theta_1 D_1 \cos \theta_2 D_2 \cos \theta_3 D_3
\Big( H^{-1} -1 \Big) I \wedge dy^1 \wedge \cdots \wedge dy^6
\end{equation}
\end{subequations}
Note that in the extremal and non-rotating case, the odd cases
$p=3,5$ can be obtained from the even cases $p=2,4$ by a T-duality
in $y^3$ and $y^5$ respectively.

For the case  $p=6$, in which the $B$-field can have the largest
rank, we also list the charges and chemical potentials associated
to the D6, D4, D2 and D0-branes respectively,
\begin{subequations}
\begin{equation}
\mu_{6} = \cos \theta_1 \cos \theta_2 \cos \theta_3 \mu \sp Q_6 =
\cos \theta_1 \cos \theta_2 \cos \theta_3 Q
\end{equation}
\begin{equation}
\label{4ch}
 \mu_{4}^{(\alpha)} = -\sin \theta_\alpha \cos \theta_\beta \cos
\theta_\gamma \mu \sp Q_4^{(\alpha)} = -\sin \theta_\alpha \cos
\theta_\beta \cos \theta_\gamma Q \sp \alpha =1 \ldots 3
\end{equation}
\begin{equation}
\label{2ch} \mu_{2}^{(\alpha)} = \cos \theta_\alpha \sin
\theta_\beta \sin \theta_\gamma  \mu \sp Q_2^{(\alpha)} =\cos
\theta_\alpha \sin \theta_\beta \sin \theta_\gamma   Q \sp \alpha
= 1\ldots 3
\end{equation}
\begin{equation}
\mu_{0} = -\sin \theta_1 \sin \theta_2 \sin \theta_3 \mu \sp Q_0 = -
\sin \theta_1 \sin \theta_2 \sin \theta_3 Q
\end{equation}
\end{subequations}
where for each $\alpha$ in \eqref{4ch}, \eqref{2ch} $\beta <
\gamma$ are both not equal to $\alpha$,
 and $\mu$ and $Q$ are the thermodynamic
quantities
\begin{equation}
\label{muandq} \mu = \tanh \alpha  \sp Q = \frac{V_p
V(S^{8-p})}{16\pi G} r_0^{7-p} (7-p) \sinh \alpha  \cosh \alpha 
\end{equation}
with $p=6$.
 For the other cases $p=2 \ldots 5$ the results follow the same
pattern, and we have in general that
\begin{equation}
\sum_{k=0}^m \sum_{\alpha} \mu_{p- 2k}^{(\alpha)}
Q_{p-2k}^{(\alpha)} = \mu Q
\end{equation}
where $\alpha$ labels the different embeddings of the
D($p-2k$)-brane into the D$p$-brane.

The near-horizon limit of these gauge potentials is most easily
expressed in terms of the T-duality invariant combinations
\begin{equation}
{\cal{A}}_{q} = (e^{-B_2} A)_{q} =A_{q} - B_2 A_{q-2} +
\frac{1}{2} B_2^2 A_{q-4} + \ldots
\end{equation}
where it is important to keep the constant parts in the gauge
potentials when present. The rescaled RR gauge potentials in the
near-horizon limit \eqref{rescal1}, \eqref{rescal2} are then
\begin{equation}
\label{rrrescal}
 {\cal{A}}_{p+1-2k} = \frac{({\cal{A}}_{p+1-2k})_{\rm
old}}{l_s^{4 + 2m -2k} } \sp k = 0 \ldots m
\end{equation}
and it can be checked that the new quantities are indeed
finite\footnote{This is provided constants in the final expression
for ${\cal{A}}$ are omitted, as also done e.g. in Ref.
\cite{Harmark:1999xt}.} in the limit $l_s \rightarrow 0$.

%---------------------------------------

\end{appendix}

\addcontentsline{toc}{section}{References}
%\bibliographystyle{JHEP}
%\bibliography{bibrot}

\providecommand{\href}[2]{#2}\begingroup\raggedright\endgroup
\end{document}